\documentclass[journal]{IEEEtran}

\IEEEoverridecommandlockouts
\usepackage{algorithm}
\usepackage{algpseudocode}
\usepackage{amsmath,amssymb,amsfonts}

\usepackage{textcomp}
\usepackage{xcolor}

\ifCLASSOPTIONcompsoc
\usepackage[caption=false,font=normalsize,labelfont=sf,textfont=sf]{subfig}
\else
\usepackage[caption=false,font=footnotesize]{subfig}
\fi

\usepackage{graphicx}
\usepackage{color}
\usepackage{xfrac}
\usepackage{amsthm}
\usepackage{relsize}
\usepackage{enumerate}
\usepackage{booktabs}
\usepackage{morefloats}
\usepackage{cite}
\usepackage{bm}
\usepackage{stfloats}
\usepackage{setspace}
\usepackage{url}
\usepackage{ulem}
\usepackage{comment}
\usepackage{flushend}
\usepackage[figuresright]{rotating}
\usepackage{multirow}
\usepackage{lscape}
\usepackage{adjustbox}
\usepackage{bm}
\usepackage{tablefootnote}
\usepackage{balance}

\newcounter{mytempeqncnt}

\begin{document}

\title{Throughput-Optimal Random Access: A Queueing-Theoretical Analysis for Learning-Based Access Design}

\author{
    Xinran Zhao and Lin Dai
    \vspace{-0.6cm}
    \thanks{The authors are with the Department of Electrical Engineering, City University of Hong Kong, Hong Kong (e-mail: xrzhao3-c@my.cityu.edu.hk; lindai@cityu.edu.hk).}
}
\maketitle

\begin{abstract}
    Random access networks have long been observed to suffer from low throughput if nodes' access strategy is not properly designed. To improve the throughput performance, learning-based approaches, with which each node learns from the observations and experience to determine its own access strategy, have shown immense potential, but are often designed empirically due to the lack of theoretical guidance. As we will demonstrate in this paper, the queueing-theoretical analysis can be leveraged as a powerful tool for optimal design of learning-based access. Specifically, based on a Multi-Armed-Bandit (MAB) framework, two random access schemes, MTOA-L with local rewards and MTOA-G with global rewards, are proposed for throughput optimization. Though both can achieve the maximum throughput of 1, they have different short-term fairness performance. Through identifying the access strategies learned via MTOA-L and MTOA-G and feeding them into the proposed unified queueing-theoretical framework, the throughput-fairness tradeoff of each is characterized and optimized by properly tuning the key parameters. The comparison of the optimal tradeoffs shows that MTOA-G is much superior to MTOA-L especially when the number of nodes is large.
\end{abstract}


\section{Introduction}\label{sec_intro}

Various random access schemes have been widely applied in communication networks. With random access, each node independently decides when to transmit. The distributed nature makes it especially appealing for next-generation wireless communication networks where Machine-to-Machine (M2M) communications is expected to play a dominant role \cite{7263367,leyva2019random}.

Despite extensive applications, it has long been observed that with random access, a network may suffer from poor access efficiency due to the lack of coordination among nodes \cite{6678832,biral2015challenges}. With the widely adopted slotted Aloha \cite{abramson1973packet}, for instance, the network throughput, which is defined as the average number of successfully transmitted data packets in the network per time slot, with the collision receiver\footnote{With the collision receiver, a packet transmission is successful if and only if there is no concurrent transmission. In that case, the network throughput is indeed the access efficiency.} is only $(1-\tfrac{1}{n})^{n-1}\approx e^{-1}$ for a large number of nodes $n$, which is far below the maximum of 1. For next-generation communication networks where random access is envisioned to facilitate a myriad of machine-type devices, it is of crucial importance to optimize the access design for maximizing the network throughput.

\subsection{Throughput-Optimal Random Access Design}\label{sec_intro-1}
\vspace{-0.05cm}
With random access, the low throughput has its root in failures caused by uncoordinated transmissions of nodes. One way to improve the throughput performance is reserving the channel for data transmissions. Specifically, each node first transmits a request to establish a connection with the receiver. Only if the request is successfully received, i.e., a connection is established, can data packets be transmitted. It was shown in \cite{8675765,10154598,10750858} that with connection-based Aloha, the network throughput can far exceed $e^{-1}$ and approach 1 as the ratio of transmission time of a data packet to a request increases.

Alternatively, the throughput performance can be improved by properly designing backoff schemes to resolve transmission failures more efficiently. It was recently shown in \cite{Aloha_capture_phenomenon} that with capture-based backoff, the network throughput of Aloha can approach 1 as well. Specifically, by setting the initial transmission probability to 1 and reducing to a small value after a certain number of failures, one node who once succeeds would have little contention from others, thus capturing the channel for a long period with a high throughput.

Note that in the above examples, nodes' access strategy is predetermined. It is the access parameters, such as the transmission probability, that are optimally tuned to maximize the network throughput. In fact, the access strategy can also be optimized, for which learning-based approaches may be helpful. From the learning perspective, random access design can be cast as a distributed sequential decision-making problem, that is, each node independently makes decisions of transmitting or not based on its own experience or observations. It can be effectively solved by reinforcement learning (RL) algorithms, where each node (agent) learns the optimal strategy for maximizing the cumulative reward.

For throughput enhancement, various RL-based random access schemes have been proposed, which can be roughly divided into two categories according to whether states are taken into account for each node's action selection. In \cite{chu2015application,8632974,silva2022throughput}, various stateless RL-based access schemes have been developed for Framed Slotted Aloha (FSA), where time slots are grouped into frames, and each node selects a slot in every frame to transmit according to the estimated mean reward of each slot. Despite the simplicity, the network throughput is found to be very sensitive to the ratio of the number of nodes to the frame size, and even drop below $e^{-1}$ when the ratio exceeds 2 \cite{silva2022throughput}. That indicates poor throughput performance for slotted Aloha, which is a special case of FSA with the frame size of 1.

On the other hand, when the local state of each node is further taken into consideration, it is often set as the past observations of the node \cite{8532121,9204466,9619960,9681886,10288545,10227309,10366306,10143731,10413951}. As storing and updating the mean rewards of all state-action pairs is intractable for a long history of observations, deep neural networks (DNNs) were utilized to map from local states to optimal actions. Despite the throughput gains, DNNs deployed at each node usually need to be trained in a centralized manner, which requires the receiver to gather a large amount of global information from all the nodes and send out the trained parameters to each node, causing heavy signaling overhead and high computational costs.

It was recently demonstrated in \cite{10423621} that without introducing states or DNN, by properly designing actions and rewards, the network throughput can reach the maximum 1 for slotted Aloha with a large number of nodes based on a simple Multi-Armed-Bandit (MAB) framework. Note that in \cite{10423621}, the reward received by each node was set to its own transmission outcome. As we will demonstrate in this paper, with such local rewards, maximizing the cumulative reward is equivalent to maximizing each node's own throughput. To optimize the total network throughput, the global reward, which is set as the aggregate outcome of all the nodes' transmissions, should be adopted. Two MAB-based throughput-optimal access (MTOA) schemes, MTOA-L and MTOA-G, are proposed based on local rewards and global rewards, respectively, and shown to both achieve the maximum network throughput of 1, but with different short-term fairness performance.
\vspace{-0.3cm}
\subsection{Optimizing the Throughput-Fairness Tradeoff: A Queueing-Theoretical Analysis for Learning-Based Random Access}\label{sec_intro-2}

For the throughput-optimal random access schemes mentioned above, the network throughput reaches the maximum of 1 when one node monopolizes the transmission, which nevertheless leads to serious unfairness among nodes. Intuitively, there exists an inherent tradeoff between the network throughput and short-term fairness\footnote{Note that here the short-term fairness refers to the fairness performance among nodes' throughput for a given time period. It should be distinguished from the long-term fairness, which can always be achieved in the symmetric case where all the nodes have the same access strategy and parameter setting. For instance, if nodes take turns to transmit but each transmits for a long period, as time goes to infinity, the network achieves the long-term fairness, but the short-term fairness could be poor.}: The longer one node can occupy the channel after each successful access, the better throughput, yet the worse short-term fairness. Instead of simply pushing the network throughput to the limit, it is of more practical significance to maximize the network throughput under a given short-term fairness requirement, or rather, to optimize the throughput-fairness tradeoff.

Both the network throughput and short-term fairness closely depend on access design and parameter setting. To optimize the throughput-fairness tradeoff, the key lies in characterizing the throughput and fairness performance of different access schemes, and understanding how the key parameters affect the tradeoff between them. For performance optimization of random access networks, the queueing-theoretical analysis has been demonstrated as a powerful tool \cite{6205590,6525472,9765641,10606277,10154598,10750858,Aloha_capture_phenomenon}. Specifically, a random access network can be regarded as a multi-queue-single-server system. By properly modeling the behavior of Head-of-Line (HOL) packets in each node's queue and establishing the fixed-point equations of steady-state probability of successful transmission of HOL packets, the service processes of nodes' queues can be characterized, based on which the network performance can further be analyzed and optimized.

The above queueing-theoretical analysis, nevertheless, requires that the access strategy of each node is given. For learning-based access design, where the access strategy is the outcome of a learning process rather than predetermined, there has been a lack of viable approaches for performance analysis. Though some parameters such as the learning rate have a significant impact on the network performance, without theoretical guidance, the parameter tuning of learning-based access schemes is inherently empirical and ad-hoc.

As we will demonstrate in this paper, the queueing-theoretical analysis can indeed be leveraged for optimal design of learning-based random access. To do that, the access strategy learned by each node should first be identified. Then the queueing-theoretical analysis can be applied to characterize the network performance with the learned access strategy, based on which the key parameters of the learning-based access schemes can further be optimized.

Specifically, the access strategy learned by each node with MTOA-L is found to be connection-free Aloha with capture-free or capture-based backoff. For MTOA-G, the access strategy is indeed connection-based Aloha with capture-free backoff. To further analyze the throughput and short-term fairness performance of MTOA-L and MTOA-G, a unified queueing-theoretical framework is established, where the design features, including connection-free or connection-based and capture-free or capture-based, are incorporated. By establishing a general Markov renewal process for HOL batches in each node's queue and deriving the steady-state probability of successful transmission of HOL batches, the network throughput and short-term fairness index are both obtained as functions of system parameters.

The analysis reveals the optimal throughput-fairness tradeoffs achieved by MTOA-L and MTOA-G as well as how to properly select the key parameters to achieve the optimal tradeoff. For instance, given that the short-term fairness index $J_T$ should be no smaller than 0.99 for $T\!=\!10^7$ time slots in a $100$-node network, the maximum achievable network throughput with MTOA-L is 0.915, to achieve which the learning rate and $Q$-value threshold should be properly tuned. With MTOA-G, on the other hand, given the same fairness constraint, the network throughput can reach the maximum of 0.998 when the number of null actions is optimally set according to the number of nodes. Significant gains can be achieved by MTOA-G thanks to the adoption of global rewards.

The remainder of the paper is organized as follows. Section \ref{sec1} presents the system model. In Section \ref{sec2}, an MAB framework is presented for throughput optimization, base on which MTOA-L and MTOA-G are proposed, and the corresponding access strategies learned by nodes are identified in Section \ref{sec3}. By establishing a unified queueing-theoretical framework in Section \ref{sec4}, the throughput-fairness tradeoffs of MTOA-L and MTOA-G are analyzed, optimized and compared in Section \ref{sec5}. Finally, conclusions are summarized in Section VII.

\section{System Model}\label{sec1}

Consider a time-slotted random access network, where $n$ nodes transmit to a single receiver. Each node is equipped with an infinite-size buffer to store the incoming data packets. All the nodes are synchronized, and can transmit only at the beginning of a time slot. The transmission of each data packet takes one time slot.

With random access, each node independently decides whether to transmit at each time slot. Due to the lack of coordination, concurrent transmissions from multiple nodes may fail. Assume the collision model at the receiver, that is, a transmission is successful if and only if there is no concurrent transmission.

For Node $i$, the output rate of its queue by time $T$ can be written as
\begin{equation}
    \lambda_{out, i}^T = \tfrac{1}{T}\sum_{t=1}^{T} N_{i,t},
    \label{eq1-1}
\end{equation}
where $N_{i,t}$ denotes the number of served (successfully transmitted) data packets of Node $i$ at time slot $t$, $i\in\mathcal{N}=\{1,2,\cdots,n\}$. With the collision model, at most one data packet can be successfully transmitted at each time slot. Therefore, $N_{i,t}\in\{0, 1\}$ for $i\in\mathcal{N}$, and $\sum_{i=1}^{n} N_{i,t}\in\{0, 1\}$. As $T\to\infty$, $\lambda_{out, i}^T$ converges to the throughput of Node $i$, i.e., the long-term average number of successfully transmitted data packets per time slot for Node $i$:
\begin{equation}
    \lambda_{out, i} = \lim_{T\to\infty} \lambda_{out, i}^T = \lim_{T\to\infty} \tfrac{1}{T}\sum_{t=1}^{T} N_{i,t},
    \label{eq1-2}
\end{equation}
$i\in\mathcal{N}$. The total network throughput can then be written as
\begin{equation}
    \hat{\lambda}_{out} = \sum_{i=1}^n \lambda_{out, i} = \lim_{T\to\infty} \tfrac{1}{T}\sum_{t=1}^{T} \sum_{i=1}^n N_{i,t}.
    \label{eq1-3}
\end{equation}
Note that to maximize the throughput of a node, its queue should be saturated, i.e., the queue is always nonempty. In this paper, we focus on the throughput optimization. Therefore, all the nodes' queues are assumed to be saturated.
\vspace{-0.3cm}
\subsection{Throughput-Optimal Random Access: A Learning Perspective}\label{sec1-1}

Both the throughput of each node $\lambda_{out, i}$ and the network throughput $\hat{\lambda}_{out}$ depend on the access design. To optimize the throughput performance for a specific access strategy, from the queueing-theoretical perspective, $\lambda_{out, i}$ and $\hat{\lambda}_{out}$ can be derived as functions of access parameters (e.g., the transmission probability of each node), and further maximized by optimally tuning access parameters. The analysis, nevertheless, requires that the access strategy is given. It is the access parameters that are optimized, rather than the access strategy.

From the learning perspective, finding the optimal access strategy of each node can be formulated as a sequential decision-making problem: In each time slot, each node independently makes a decision of transmitting or not based on its observations and past experience. To solve sequential decision-making problems, RL has been widely adopted. In the RL framework, an agent interacts with the environment by observing the state, taking an action, and receiving a reward at each step. The goal of the agent is to find the optimal strategy for maximizing the mean cumulative reward.

To utilize RL algorithms for access strategy optimization, the key lies in the design of states, actions and rewards. Generally speaking, states depend on the network status that can be observed by each node, actions are usually related to transmission decisions, and rewards should be set according to the goal. It has been demonstrated in \cite{10423621} that based on a simple MAB framework, by properly designing actions and rewards, the network throughput can approach the maximum 1. It will be further shown in Section \ref{sec2} that by setting the reward as each node's own transmission outcome or the global outcome of all the nodes' transmissions, maximizing the cumulative reward becomes equivalent to maximizing each node's own throughput or maximizing the network throughput, respectively. Both MTOA-L and MTOA-G can achieve the maximum network throughput of 1, though with different throughput-fairness tradeoffs.

\vspace{-0.3cm}
\subsection{Tradeoff Between Network Throughput and Short-Term Fairness}\label{sec1-2}

It can be seen from (\ref{eq1-2}) and (\ref{eq1-3}) that if only one node transmits all the time, both the node's throughput and the network throughput can reach the maximum of 1. It nevertheless leads to serious unfairness as the other $n-1$ nodes cannot transmit.

To evaluate the fairness performance, we introduce the Jain's fairness index \cite{jain1984quantitative} as
\begin{equation}
    J_T \triangleq \frac{(\sum_{i=1}^{n}\lambda_{out,i}^T)^2}{n\cdot\sum_{i=1}^{n}(\lambda_{out,i}^T)^2},
    \label{eq1-4}
\end{equation}
for a given period of time $T$. In this paper, $J_T$ is referred to as the short-term fairness index because it is evaluated for a given time period $T$. Note that a larger $J_T$ indicates better fairness. If all the nodes' queues have identical output rates, $J_T=1$. If only one node transmits by time $T$, then $J_T=\tfrac{1}{n}\ll 1$ for a large number of nodes $n$.

Intuitively, the longer one node occupies the channel and monopolizes the transmission, the higher network throughput, yet the worse short-term fairness performance. Apparently, the throughput-fairness tradeoff is crucially determined by the access strategy and access parameter setting. To understand how to optimize the throughput-fairness tradeoffs of the proposed MTOA-L and MTOA-G, we will first identify the access strategies learned via MTOA-L and MTOA-G in Section \ref{sec3}, and then establish a unified queueing-theoretical framework for analyzing the network throughput and short-term fairness performance in Section \ref{sec4}. The optimal throughput-fairness tradeoffs for MTOA-L and MTOA-G are characterized and compared in Section \ref{sec5}.

\vspace{-0.2cm}
\section{MAB-Based Throughput-Optimal Random Access}\label{sec2}

In this section, an MAB framework for throughput optimization will be presented, based on which two MAB-based throughput-optimal random access schemes, MTOA-L and MTOA-G, will be proposed.
\vspace{-0.3cm}
\subsection{MAB Framework for Throughput Optimization}\label{sec2-1}

In an MAB framework, each node takes an action at each time slot, and receives a reward based on the outcome of its action. The objective is to find the optimal strategy to maximize the expected cumulative reward. In the following, we will demonstrate how to properly set the rewards and actions for optimizing the throughput performance of a random access network.

\subsubsection{Local Rewards and Global Rewards}\label{sec2-1-1}

Let $r_t^i$ denote the reward received by Node $i$ at time slot $t$, $i\in\mathcal{N}$. If $r_t^i$ is set as
\begin{equation}
    r_t^i = N_{i,t},
    \label{eq2-1-2}
\end{equation}
then $r_t^i=1$ if Node $i$ has a successful transmission, and $r_t^i=0$ otherwise, $i\in\mathcal{N}$. For each node, if it transmits and receives an acknowledgement (ACK) indicating its successful transmission, the reward is 1; otherwise, the reward is 0. By combining (\ref{eq2-1-2}) with (\ref{eq1-2}), we have $\lim_{T\to\infty} \tfrac{1}{T}\sum_{t=1}^{T} r_t^i = \lambda_{out,i}$, indicating that maximizing the cumulative reward is equivalent to maximizing Node $i$'s throughput $\lambda_{out,i}$.

On the other hand, if $r_t^i$ is set as
\begin{equation}
    r_t^i = \sum_{j=1}^{n} N_{j,t},
    \label{eq2-1-4}
\end{equation}
then $r_t^i=1$ if the network has a successful transmission (which may not be from Node $i$), and $r_t^i=0$ otherwise, $i\in\mathcal{N}$. In other words, when the receiver broadcasts an ACK indicating the successful transmission of a node, all the nodes receive a reward of $1$; otherwise, they receive a reward of 0. By combining (\ref{eq2-1-4}) with (\ref{eq1-3}), we have $\lim_{T\to\infty} \tfrac{1}{T}\sum_{t=1}^{T} r_t^i = \hat{\lambda}_{out}$, indicating that maximizing the cumulative reward is equivalent to maximizing the network throughput $\hat{\lambda}_{out}$.

We can see that in the former case, each node receives a reward only based on the outcome of its own transmission, while in the latter case, the reward is based on the outcome of all the transmissions in the network. We refer to the rewards given by (\ref{eq2-1-2}) and (\ref{eq2-1-4}) as local reward and global reward, respectively. In both cases, the objective is to optimize the throughput performance, i.e., maximizing each node's throughput based on local rewards or maximizing the network throughput based on global rewards. The corresponding MAB schemes are referred to as MTOA-L and MTOA-G, respectively.

\subsubsection{Actions}\label{sec2-1-2}

Let $A_t^i$ denote the action taken by Node $i$ at time slot $t$, $i\in\mathcal{N}$. $A_t^i$ is chosen from the action set $\mathcal{A}$, which is defined as $\mathcal{A}=\{0,1,\cdots,L\}$. $A_t^i=0$ indicates that Node $i$ transmits at time slot $t$, and $A_t^i=l\in\{1,\cdots,L\}$ indicates that Node $i$ does not transmit at time slot $t$, where $L$ is the number of null actions. At each time slot, each node selects the action with the highest estimated mean reward, i.e., $A_t^i = \arg \max_{a\in\mathcal{A}} Q_{t}(i, a)$, where $Q_{t}(i, a)$ denotes the mean reward of action $a$ estimated by Node $i$ at time slot $t$, i.e., the $Q$ value, which is updated as
\begin{equation}
    Q_{t+1}(i, A_t^i) = Q_{t}(i, A_t^i) + \alpha(r_t^i - Q_{t}(i, A_t^i)),
    \label{eq2-1-6}
\end{equation}
after Node $i$ takes the action $A_t^i$ and receives the reward $r_t^i$, $i\in\mathcal{N}$. In (\ref{eq2-1-6}), $\alpha\in(0,1]$ denotes the learning rate. All $Q$ values are initialized to zero. If there are multiple actions with the same highest $Q$ value, a node will choose one of them with equal probability.

In the following sections, we will present more details on MTOA-L and MTOA-G, and evaluate their performance.

\subsection{MTOA-L}\label{sec2-2}

With MTOA-L, only the node who succeeds receives a local reward of 1. The $Q$ values for the null actions are always zero. With the learning rate $\alpha<1$, once a node succeeds, its $Q$ value for the transmission action would always be positive according to (\ref{eq2-1-6}), and thus it would transmit forever. When the number of null actions $L\to \infty$, the other nodes never transmit, and thus the node's transmissions could always succeed, leading to a network throughput of 1. Nevertheless, the other nodes' throughput is zero, indicating poor fairness performance.

To address the unfairness problem in MTOA-L, we introduce a threshold $Q_{th}$, below which a positive $Q$ value is reset to zero. As a node's positive $Q$ value is reduced every time its transmission fails, with a positive threshold $Q_{th}$, a node who once succeeds would reset its $Q$ value of the transmission action to zero if one of its packets experiences a certain number of failures. By preventing a node from transmitting forever, the fairness performance can be improved. The detailed procedure of MTOA-L is presented in Algorithm \ref{alg1}.\footnote{Note that different from MTOA-L where the $Q$-value threshold is adopted to tackle the unfairness issue, in \cite{10423621}, an exploration probability $\varepsilon$ was introduced. Specifically, at each time slot, with probability $1-\varepsilon$, each node chooses the action with the highest $Q$ value. With probability $\varepsilon$, a node randomly chooses an action from the action set. Despite the improvement in the short-term fairness, the network throughput is significantly degraded. Simulation results show that given the same fairness requirement, MTOA-L achieves a higher network throughput than the one proposed in \cite{10423621}.}

\algnewcommand{\Initialize}[1]{%
  \State \textbf{Initialize:}
  \Statex \hspace*{\algorithmicindent}\parbox[t]{.8\linewidth}{\raggedright #1}
}
\begin{algorithm}[t]
    \caption{MTOA-L}\label{alg1}
    \begin{algorithmic}
        \Initialize{$Q(i,a)=0$ for $\forall$ $i\in\mathcal{N}$ and $\forall$ $a\in\mathcal{A}$.}
        \For{$t = 1,2,\cdots,T$}
            \For{Node $i = 1,2,\cdots,n$}
                \State $A_t^i = \arg \max_{a\in\mathcal{A}} Q(i, a)$.
                \State Execute the action $A_t^i$ and receive the local reward $r_t^i = N_{i,t}$.
                \State $Q(i, A_t^i) = Q(i, A_t^i) + \alpha(r_t^i - Q(i, A_t^i))$.
                \If{$Q(i, A_t^i) \leq Q_{th}$}
                \State $Q(i, A_t^i)=0$.
                \EndIf
            \EndFor
        \EndFor
    \end{algorithmic}
\end{algorithm}

\begin{figure*}[t]
    \centering
    \subfloat[]{
        \includegraphics[width=0.32\textwidth, height=0.19\textwidth]{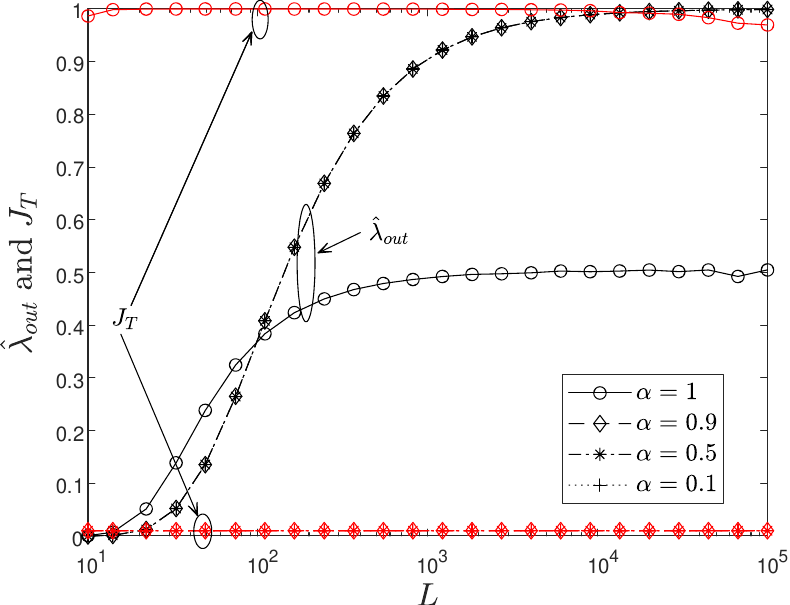}
        \label{fig2-2-1-1}
    }
    \subfloat[]{
        \includegraphics[width=0.32\textwidth, height=0.19\textwidth]{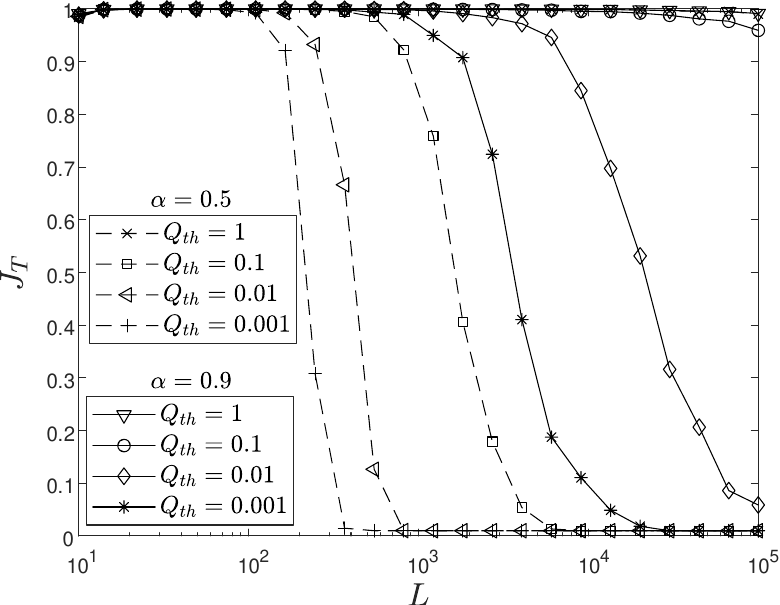}
        \label{fig2-2-1-2}
    }
    \subfloat[]{
        \includegraphics[width=0.32\textwidth, height=0.19\textwidth]{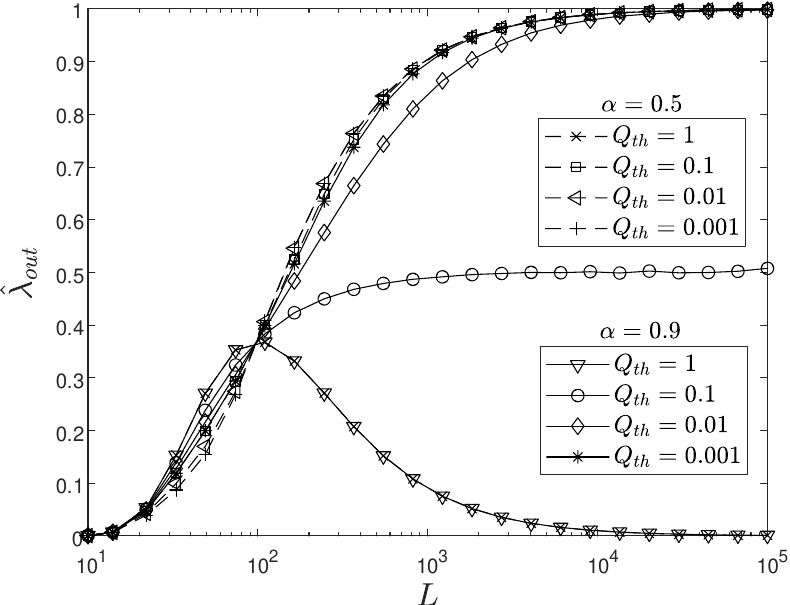}
        \label{fig2-2-1-3}
    }
    \vspace{-0.15cm}\caption{Simulated network throughput $\hat{\lambda}_{out}$ and short-term fairness index $J_T$ with MTOA-L versus the number of null actions $L$. $n=100$. $T=10^7$ slots. (a) $Q_{th}=0$. $\alpha\in\{0.1,0.5,0.9,1\}$. (b)-(c) $Q_{th}\in\{0.001,0.01,0.1,1\}$. $\alpha\in\{0.5,0.9\}$.}
    \label{fig2-2-1}
\end{figure*}

Fig. \ref{fig2-2-1} presents the simulated network throughput $\hat{\lambda}_{out}$ and short-term fairness index $J_T$ with MTOA-L. It can be seen from Fig. \ref{fig2-2-1-1} that with the $Q$-value threshold $Q_{th}=0$ and learning rate $\alpha<1$, $\hat{\lambda}_{out}$ approaches 1 when the number of null actions $L$ is large, while $J_T$ is always equal to $\tfrac{1}{n}=0.01$ as only one node has non-zero throughput. With a positive $Q_{th}$ and $\alpha<1$, on the other hand, the fairness performance is significantly improved, especially when $L$ is small, as shown in Fig. \ref{fig2-2-1-2}.

Further note from Fig. \ref{fig2-2-1-2}-\ref{fig2-2-1-3} that when $0<Q_{th}<1$, despite the improvement in fairness, a small number of null actions $L$ leads to a low network throughput $\hat{\lambda}_{out}$. By increasing $L$, $\hat{\lambda}_{out}$ is enlarged, though at the cost of a smaller short-term fairness index $J_T$. It suggests that the number of null actions $L$ determines a fundamental tradeoff between the network throughput and short-term fairness. 

The throughput-fairness tradeoff depends on many factors such as the $Q$-value threshold $Q_{th}$ and learning rate $\alpha$. As we can see in Fig. \ref{fig2-2-1-2}, increasing $Q_{th}$ and $\alpha$ can improve the short-term fairness. Yet the throughput performance may sharply decline if $Q_{th}$ is too large, e.g., $Q_{th}=1$, as shown in Fig. \ref{fig2-2-1-3}. It implies that to optimize the throughput-fairness tradeoff, the $Q$-value threshold $Q_{th}$ and learning rate $\alpha$ should be properly chosen.

\begin{figure*}[t]
    \vspace{-0.5cm}
    \centering
    \subfloat[]{
        \includegraphics[width=0.32\textwidth, height=0.19\textwidth]{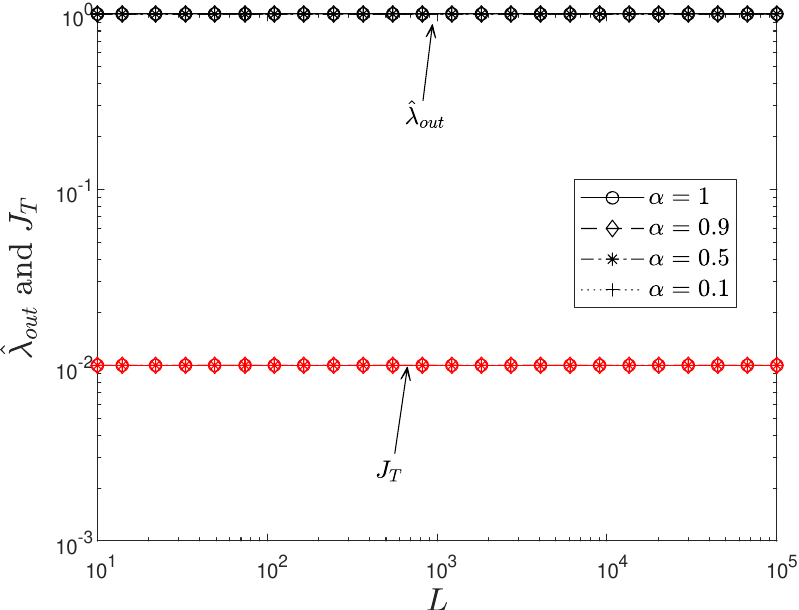}
        \label{fig2-3-1-1}
    }
    \subfloat[]{
        \includegraphics[width=0.32\textwidth, height=0.19\textwidth]{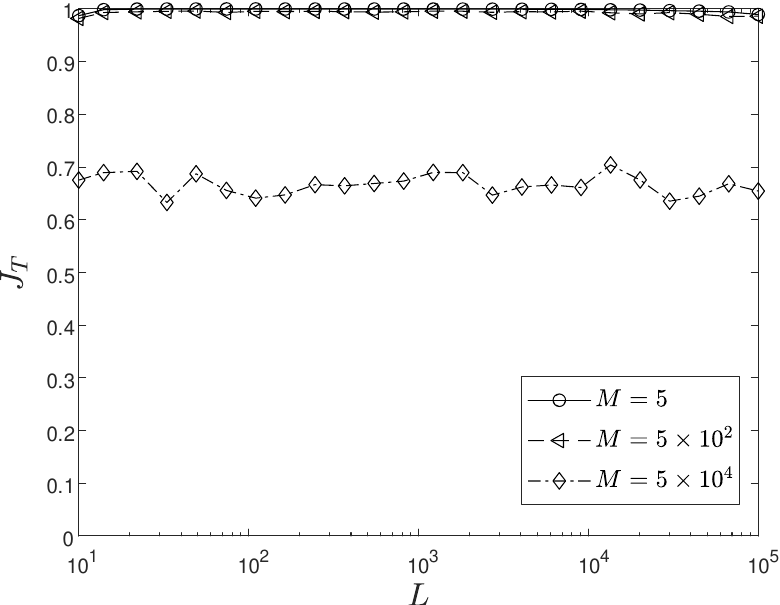}
        \label{fig2-3-1-2}
    }
    \subfloat[]{
        \includegraphics[width=0.32\textwidth, height=0.19\textwidth]{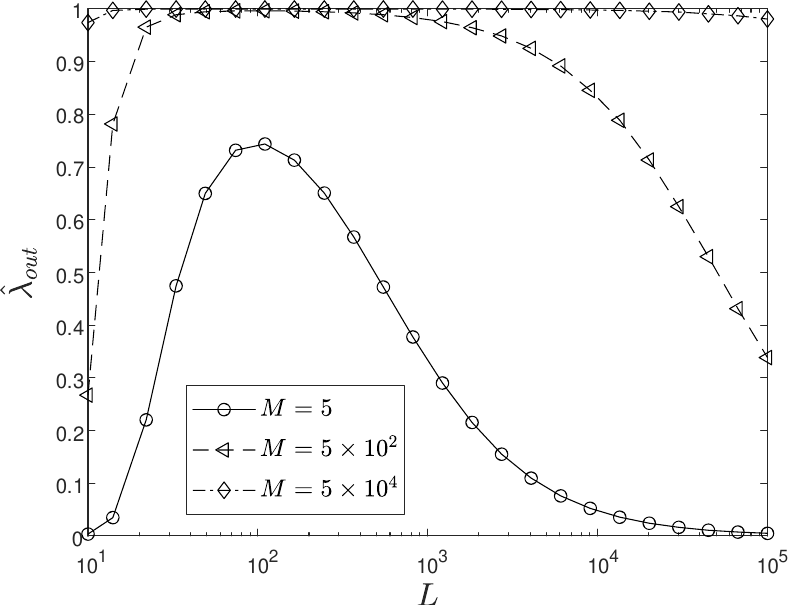}
        \label{fig2-3-1-3}
    }
    \vspace{-0.15cm}\caption{Simulated network throughput $\hat{\lambda}_{out}$ and short-term fairness index $J_T$ with MTOA-G versus the number of null actions $L$. $n=100$. $T=10^7$ slots. (a) $M=\infty$. $\alpha\in\{0.1,0.5,0.9,1\}$. (b)-(c) $\alpha=0.9$. $M\in\{5,5\times 10^2,5\times 10^4\}$.}
    \label{fig2-3-1}
    \vspace{-0.2cm}
\end{figure*}
\vspace{-0.3cm}
\subsection{MTOA-G}\label{sec2-3}

\begin{algorithm}[t]
    \caption{MTOA-G}\label{alg2}
    \begin{algorithmic}
        \Initialize{$Q(i,a)=0$ for $\forall$ $i\in\mathcal{N}$ and $\forall$ $a\in\mathcal{A}$, and $W_i=0$ for $\forall$ $i\in\mathcal{N}$.}
        \For{$t = 1,2,\cdots,T$}
            \For{Node $i = 1,2,\cdots,n$}
                \State $A_t^i = \arg \max_{a\in\mathcal{A}} Q(i, a)$.
                \State Execute the action $A_t^i$ and receive the global reward $r_t^i = \sum_{j=1}^{n} N_{j,t}$.
                \State $Q(i, A_t^i) = Q(i, A_t^i) + \alpha(r_t^i - Q(i, A_t^i))$.
                \If{$Q(i, A_t^i)>0$}
                    \State $W_i = W_i + 1$.
                    \If{$W_i == M$}
                        \State $W_i = 0$.
                        \State $Q(i, A_t^i)=0$.
                    \EndIf
                \EndIf
            \EndFor
        \EndFor
    \end{algorithmic}
\end{algorithm}

With MTOA-G, if only one node transmits and succeeds, all the nodes receive a global reward of 1. For the node who transmits, the $Q$ value of the transmission action becomes positive; for the other nodes, the $Q$ values of their chosen null actions become positive. As a result, a node that once succeeds would transmit forever and never fail as the other nodes never interrupt. Therefore, MTOA-G can also achieve the network throughput of 1, though with severe unfairness.

Note that the unfairness problem in MTOA-G cannot be addressed by introducing the $Q$-value threshold used in MTOA-L. That is because with MTOA-G, once a node succeeds, all the nodes always receive a global reward of 1, and their positive $Q$ values never decrease. To improve the fairness performance of MTOA-G, we introduce a $Q$-value reset window size $M$ instead. Specifically, each node counts the number of time slots during which its $Q$ value is positive, and resets the $Q$ value to zero when the counter reaches $M$. Intuitively, a smaller $M$ leads to better fairness performance as each node occupies the channel for shorter time. The detailed procedure of MTOA-G is presented in Algorithm \ref{alg2}.

Fig. \ref{fig2-3-1} illustrates the simulated network throughput $\hat{\lambda}_{out}$ and short-term fairness index $J_T$ with MTOA-G under different settings of $Q$-value reset window size $M$. We can see from Fig. \ref{fig2-3-1-1} that with $M=\infty$, in which case the positive $Q$ values are never reset to zero, the network throughput of 1 can always be achieved, yet the fairness performance is poor regardless of the number of null actions $L$ and learning rate $\alpha$.

When $M<\infty$, it can be seen from Fig. \ref{fig2-3-1-2}-\ref{fig2-3-1-3} that a decrease in $M$ leads to a larger short-term fairness index $J_T$, yet a smaller network throughput $\hat{\lambda}_{out}$. It indicates that $M$ determines a tradeoff between the throughput and fairness performance of MTOA-G. Moreover, for given $M$, the short-term fairness index $J_T$ is observed to be insensitive to the number of null actions $L$, while the network throughput $\hat{\lambda}_{out}$ is sensitive to $L$, especially when $M$ is small. It suggests that $L$ should be carefully selected for achieving the optimal throughput-fairness tradeoff.
\vspace{-0.2cm}
\subsection{Summary}\label{sec2-4}

To conclude, for both MTOA-L and MTOA-G, the maximum network throughput of 1 can be achieved, though with severe unfairness as only one node's throughput is non-zero. By introducing a positive $Q$-value threshold $Q_{th}$ for MTOA-L and a finite $Q$-value reset window size $M$ for MTOA-G, the fairness performance can be improved, especially when $Q_{th}$ is large or $M$ is small. A too large $Q_{th}$ or too small $M$, nevertheless, may lead to degradation in the throughput performance. In addition to $Q_{th}$ and $M$, the network throughput and short-term fairness are also dependent on other parameters such as the number of null actions $L$ and learning rate $\alpha$.

The preceding results indicate that for both MTOA-L and MTOA-G, there exists a tradeoff between the network throughput and short-term fairness performance, which depends on many factors. It is important to understand how to optimize the throughput-fairness tradeoff by properly tuning the parameters. Yet the MAB framework does not offer a viable approach for analyzing the throughput and fairness performance of MTOA-L and MTOA-G. As we will demonstrate in the following sections, we can resort to the queueing-theoretical analysis for characterizing and further optimizing the throughput-fairness tradeoff of the learning-based MTOA-L and MTOA-G. To leverage the queueing-theoretical analysis, we need to first know what access strategies nodes learn via MTOA-L and MTOA-G, which will be identified in the next section.

\section{Access Strategies Learned via MTOA-L and MTOA-G}\label{sec3}

In this section, we will look into the access strategies learned by nodes via MTOA-L and MTOA-G, and extract the design features of the learned access strategies.
\vspace{-0.2cm}
\subsection{Access Strategy of MTOA-L}\label{sec3-1}

With MTOA-L, if a node (say Node $i$, $i\in\mathcal{N}$) transmits and succeeds, it receives a local reward of 1, while the other nodes receive local rewards of 0. Then Node $i$ transmits its HOL data packet with probability 1 as its $Q$ value for the transmission action is positive, and Node $j$ selects the transmission action with probability $\tfrac{1}{L+1}$ as its $Q$ values for both the transmission action and $L$ null actions are zero, $j\in\mathcal{N}\backslash \{i\}$.

For Node $i$, its positive $Q$ value of the transmission action, $Q_{t}(i,0)$, would be reset to zero when it drops below the threshold $Q_{th}$. Note from (\ref{eq2-1-6}) that $Q_{t}(i,0)$ is decreased to $(1-\alpha)Q_{t}(i,0)$ if Node $i$'s transmission is interrupted by the other nodes'. As a consequence, if one of Node $i$'s HOL packets experiences a certain number of failures such that $Q_{t}(i,0)$ drops below $Q_{th}$, its transmission probability would be reduced from 1 to $\tfrac{1}{L+1}$.

For each data packet, let $Q_0$ denote its associated node's $Q$ value of the transmission action when it becomes a HOL packet, and $n_C$ denote the number of failures experienced by it at which the $Q$ value of the transmission action drops below $Q_{th}$. Apparently, $n_C$ is determined by $\alpha$, $Q_{th}$ and $Q_0$. Specifically, with learning rate $\alpha<1$ and $Q$-value threshold $Q_{th}<\alpha$, we have $n_C = \Bigl\lceil \log_{1-\alpha}\tfrac{Q_{th}}{Q_0} \Bigr\rceil$. With $\alpha=1$ and $Q_{th}< \alpha$, as a node resets its $Q$ value of the transmission action to zero once its transmission fails, we have $n_C=1$. With $Q_{th} \geq \alpha$, we have $n_C=0$ by noting that even though a node succeeds and updates its $Q$ value of the transmission action to $\alpha$, it would immediately reset the positive $Q$ value to zero.

To summarize, the access strategy learned via MTOA-L can be described as follows:

\textit{\textbf{MTOA-L:} Each node transmits its HOL packet at each time slot with probability $q_k$, where $q_k$ is adjusted according to the number of failures experienced by the HOL packet as
\vspace{0.1cm}
\begin{equation}
    q_k =
    \begin{cases}
        1 & 0\leq k \leq n_C-1 \\
        \tfrac{1}{L+1} & k\geq n_C,
    \end{cases}
    \label{eq3-1-1}
\end{equation}
where
\vspace{0.1cm}
\begin{equation}
    n_C =
    \begin{cases}
        0 & Q_{th}\geq \alpha \\
        1 & Q_{th}< \alpha = 1 \\
        \Bigl\lceil \log_{1-\alpha}\frac{Q_{th}}{Q_0} \Bigr\rceil & Q_{th}< \alpha < 1.
    \end{cases}
    \label{eq3-1-2}
\end{equation}
}
\vspace{-0.45cm}
\subsection{Access Strategy of MTOA-G}\label{sec3-2}

With MTOA-G, if a node (say Node $i$, $i\in\mathcal{N}$) chooses the transmission action and the other nodes choose null actions, all the nodes receive a global reward of 1, and thus the $Q$ values for Node $i$'s transmission action and the other nodes' chosen null actions become positive. Then Node $i$ transmits its HOL data packet with probability 1, while all the other nodes do not transmit. Note that in MTOA-G, a $Q$ value would be reset to zero if it remains positive for $M$ time slots. As a result, after $M$ consecutive successful transmissions from Node $i$, all the nodes reset their positive $Q$ values to zero, and contend for access with transmission probability $\tfrac{1}{L+1}$.

To summarize, the access strategy learned via MTOA-G can be described as follows:

\textit{\textbf{MTOA-G:} Each node transmits its HOL packet at each time slot with probability $\tfrac{1}{L+1}$. If a node makes a successful transmission at a time slot, it would transmit $M-1$ data packets in the following $M-1$ time slots, during which the other nodes do not transmit. After the successful transmission of $M$ data packets, all the nodes re-contend with transmission probability $\tfrac{1}{L+1}$.}
\vspace{-0.2cm}
\subsection{Categorization of Random Access Schemes}\label{sec3-3}

Both MTOA-L and MTOA-G are essentially random access schemes. In this section, we will categorize the random access schemes, and identify the design features of the access strategies learned via MTOA-L and MTOA-G.

\subsubsection{Sensing-Free Access versus Sensing-Based Access}\label{se3-3-1}

According to whether nodes have the sensing capability, random access protocols can be divided into two categories: sensing-free Aloha and sensing-based Carrier Sense Multiple Access (CSMA). Different from Aloha where sensing is not required, with CSMA, if a node has packets to transmit, it needs to sense the channel first. Only when the channel is sensed idle can it transmit with a certain probability at the next time slot.

\subsubsection{Connection-Free Access versus Connection-Based Access}\label{se3-3-2}

According to whether every data packet contends for access, random access protocols can also be categorized into connection-based ones and connection-free ones. Specifically, with connection-based random access, each node transmits a data packet for contention first.\footnote{Note that in practice, with connection-based random access, each node sends a request instead of a data packet for establishing a connection with the receiver. Examples include the 4-step random access procedure in 5G cellular systems \cite{3gpp.38.321} and the request-to-send/clear-to-send (RTS/CTS) access mechanism in WiFi networks \cite{ieee_802_11_MAC}. As the requests carry no data, extra overhead is introduced, which reduces the maximum data throughput that can be achieved \cite{10750858}.} If an ACK that indicates the successful transmission is broadcast by the receiver, i.e., a connection is established, then $M-1$ data packets can be transmitted over the reserved channel. Here $M$ is the total number of data packets that can be consecutively transmitted by a node every time it succeeds in contention. With connection-free random access, on the other hand, every data packet has to contend, i.e., $M=1$.

\subsubsection{Capture-Free Backoff versus Capture-Based Backoff}\label{sec3-3-3}

Backoff is a key component of random access protocols for resolving transmission failures. Specifically, each node adjusts its transmission probability according to the number of failures experienced by its HOL packet. Let $q_k$ denote the transmission probability of each node after the $k$-th failure of its HOL packet, $k=0,1,\cdots$. Typically, $\{q_k\}$ is a non-increasing sequence, i.e., $q_{k+1}\leq q_k$ for $k=0,1,\cdots$.

If one node transmits with probability 1 while the other nodes transmit with a small probability, the node has little contention, and thus may capture the channel and transmit a stream of data packets. Whether the above channel-capture phenomenon \cite{386597,Aloha_capture_phenomenon} occurs crucially depends on the backoff design. For instance, if the transmission probability for each HOL packet after its $k$-th failure is $q_k=1$ for $0\leq k\leq n_C-1$ and $q_k \ll 1$ for $k\geq n_C$, then with $n_C=0$, the channel-capture phenomenon would not occur since each node always has a small $q_k$ for $k\geq 0$ and cannot continuously transmit its data packets. Only with $n_C\geq 1$ may the channel-capture phenomenon be induced. In general, let $n_C$ denote the number of capture states. Based on the value of $n_C$, the backoff schemes can be divided into two categories, capture-free with $n_C=0$ and capture-based with $n_C\geq 1$.
\vspace{-0.1cm}
\subsection{Summary}\label{sec3-4}

Let us look back at the access strategies learned via MTOA-L and MTOA-G. Clearly, both are sensing-free Aloha schemes as nodes do not sense before their transmissions. Furthermore, the one with MTOA-G is connection-based as a node who once succeeds can exclusively transmit $M-1$ data packets without contention from the other nodes, while the one with MTOA-L is connection-free as every data packet has to contend. As for the backoff strategy, with MTOA-L, note from (\ref{eq3-1-2}) that the number of capture states $n_C$ depends on the learning rate $\alpha$ and $Q$-value threshold $Q_{th}$. With $Q_{th}\geq \alpha$, a capture-free backoff strategy would be learned. Otherwise, the backoff strategy is capture-based. With MTOA-G, the learned backoff strategy is capture-free as each HOL packet involved in contention is transmitted with probability $\tfrac{1}{L+1}$ for $L\geq 1$.

\begin{table}[t]
    \caption{Design Features of MTOA-L and MTOA-G}
    \label{table1}
    \resizebox{\columnwidth}{!}{%
    \begin{tabular}{|c|c|c|}
    \hline
                          & MTOA-L                                                                                                      & MTOA-G           \\ \hline
    Sensing-Free/Based    & Sensing-Free                                                                                                & Sensing-Free     \\ \hline
    Connection-Free/Based & Connection-Free                                                                                             & Connection-Based \\ \hline
    Capture-Free/Based    & \begin{tabular}[c]{@{}c@{}}$Q_{th}\geq \alpha$: Capture-Free\\ $Q_{th}< \alpha$: Capture-Based\end{tabular} & Capture-Free     \\ \hline
    \end{tabular}%
    }
\end{table}

The design features of the access strategies with MTOA-L and MTOA-G are summarized in Table \ref{table1}. To analyze their performance, in the next section, a unified queueing-theoretical framework that incorporates both connection-free or connection-based and capture-free or capture-based is established for Aloha networks, based on which the network throughput and short-term fairness are characterized.
\vspace{-0.1cm}
\section{A Unified Queueing-Theoretical Framework for Random Access}\label{sec4}

Essentially, an $n$-node random access network can be regarded as a multi-queue-single-server system, the performance of which is crucially determined by the service processes of nodes' queues. In general, data packets in each queue are served in a batch of $M\geq 1$ packets. Specifically, in a batch, only the first packet contends for access, and the remaining $M-1$ packets are transmitted over the reserved channel after the successful transmission of the first packet. Note that $M>1$ for connection-based access. For connection-free access, on the other hand, $M=1$ as every data packet has to contend for access.

For each node's queue, the key to characterizing its service process lies in properly modeling the behavior of its HOL batches. As we will demonstrate in the following subsection, the behavior of each HOL batch can be characterized by a general Markov renewal process, in which the holding time of each state and the state transition probabilities are dependent on the access design features such as connection-based or connection-free and capture-based or capture-free.
\vspace{-0.2cm}
\subsection{State Characterization of HOL Batches}\label{sec4-1}

\begin{figure}[t]
    \centering
    \includegraphics[width=0.98\columnwidth, height=0.25\columnwidth]{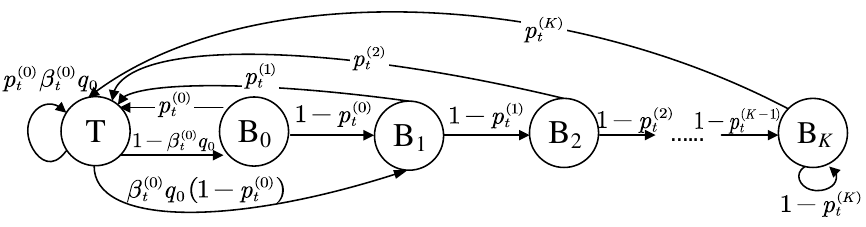}
    \caption{Embedded Markov chain $\mathbf{X}$ of the state transition process of an individual HOL batch in the Aloha network.}
    \label{fig3-2-2}
\end{figure}

For the Aloha network, the behavior of each HOL batch can be modeled by a discrete-time Markov renewal process $(\mathbf{X}, \mathbf{V})=\{(X_i, V_i),\;i=0,1,...\}$, where $X_i$ denotes the state of a tagged HOL batch at the $i$-th transition and $V_i$ denotes the epoch at which the $i$-th transition occurs. Fig. \ref{fig3-2-2} illustrates the embedded Markov chain $\mathbf{X}=\{X_i\}$. For each HOL batch, there are 1) one successful-transmission state, State T, and 2) $K+1$ failed-transmission/backoff states, State B$_k$, $k\in\{0,1,\cdots,K\}$. Here $K$ is the cutoff phase, that is, if the number of failures experienced by a HOL batch $k\geq K$, the transmission probability $q_k$ remains unchanged with $q_k=q_K$. It is usually adopted to prevent the transmission probability from being excessively small.

For a fresh HOL batch, if the first packet is transmitted and succeeds, it would stay in State T for $M$ time slots. Otherwise, it would move to State B$_0$ if the transmission of the first packet is suspended, or State B$_1$ if the transmission fails. For a State-B$_k$ HOL batch, its first packet is transmitted with probability $q_k$ given that the channel is not reserved, $k\in\{0,1,\cdots,K\}$. If its first packet is successfully transmitted, it shifts to State T; otherwise, it moves to State B$_{\min\{k+1,K\}}$.

For each HOL batch whose first packet is at the backoff stage $k$, let $\beta_t^{(k)}$ denote the probability that the channel is not reserved at time slot $t$, and $p_t^{(k)}$ denote the probability that the transmission of its first packet is successful at time slot $t$ given that the channel is not reserved. The steady-state probability distribution of the embedded Markov chain $\mathbf{X}$ in Fig. \ref{fig3-2-2} can be obtained as
\begin{equation}
    \begin{split}
        \pi_T \!=\! \tfrac{1}{2-\beta^{(0)}q_0+\sum_{k=1}^{K-1}\prod_{l=0}^{k-1}(1-p^{(l)})+\tfrac{1}{p^{(K)}}\prod_{l=0}^{K-1}(1-p^{(l)})},\\
        \pi_{B_k} \!=\! 
        \begin{cases}
            (1-\beta^{(0)}q_0)\pi_T & k=0 \\
            \pi_T\prod_{l=0}^{k-1}(1-p^{(l)}) & 1 \leq k \leq K-1 \\
            \tfrac{\pi_T}{p^{(K)}}\prod_{l=0}^{K-1}(1-p^{(l)}) & k=K,
        \end{cases}
    \end{split}
    \label{eq3-2-1-1}
\end{equation}
where $\beta^{(k)} = \lim \limits_{t \to \infty} \beta_t^{(k)}$ and $p^{(k)} = \lim \limits_{t \to \infty} p_t^{(k)}$. The mean holding time of each state is given by
\begin{equation}
    \tau_T = M, \hspace{0.8cm} \tau_{B_k} = \tfrac{1}{\beta^{(k)}q_k},\;\;\;0 \leq k \leq K.
    \label{eq3-2-1-2}
\end{equation}

The limiting state probabilities of the Markov renewal process $(\mathbf{X}, \mathbf{V})$ are given by
\begin{equation}
    \tilde{\pi}_u = \frac{\pi_u\tau_u}{\Sigma_{v\in\mathcal{S}}\pi_v\tau_v},\,\,\,u\in\mathcal{S},
    \label{eq3-2-1-3}
\end{equation}
where $\mathcal{S}=\{\text{T},\text{B}_0,\cdots,\text{B}_K\}$ is the state space of embedded Markov chain $\mathbf{X}$. By combining (\ref{eq3-2-1-1})-(\ref{eq3-2-1-3}), the limiting state probabilities $\{\tilde{\pi}_u\}_{u\in\mathcal{S}}$ of the Markov renewal process $(\mathbf{X}, \mathbf{V})$ can then be obtained.

It can be seen from (\ref{eq3-2-1-1})-(\ref{eq3-2-1-3}) that the limiting state probabilities $\{\tilde{\pi}_u\}_{u\in\mathcal{S}}$ are determined by the steady-state probability of the channel being not reserved $\beta^{(k)}$ and the steady-state probability of successful transmission given that the channel is not reserved $p^{(k)}$, which will be derived in the following section.
\vspace{-0.1cm}
\subsection{Steady-State Probabilities $\beta^{(k)}$ and $p^{(k)}$}\label{sec4-2}

Whether the steady-state probability of the channel being not reserved, $\beta^{(k)}$, and steady-state probability of successful transmission given that the channel is not reserved, $p^{(k)}$, depend on the backoff stage $k$ of each HOL batch is determined by the backoff design. In the literature, $p^{(k)}$ is usually assumed to be independent of the backoff stage, i.e., $p^{(k)}=p$ for all $0\leq k \leq K$. It was pointed out in \cite{Aloha_capture_phenomenon} that when the channel-capture phenomenon is induced by capture-based backoff, this state independence assumption does not hold any more. Specifically, if a node has a fresh HOL packet and transmits with probability 1, all the other nodes' HOL packets cannot succeed, and must be pushed to deep backoff stages with small transmission probabilities. Thus, a fresh HOL packet has a higher chance of success than those in deep backoff stages.

Note that \cite{Aloha_capture_phenomenon} considers connection-free Aloha. In general, for both connection-free and connection-based Aloha, the states of each HOL batch can be divided into two categories: capture and non-capture. Similar to \cite{Aloha_capture_phenomenon}, define a state as a \textit{Capture State} if it satisfies the following two conditions:
\begin{itemize}
    \item If a HOL batch is in a capture state, then the probability of successful transmission of other HOL batches is zero;
    \item If a HOL batch is in a capture state, then no other HOL batches can be in a capture state.
\end{itemize}
Otherwise, it is a non-capture state.

Let $\mathcal{S}_C$ and $\mathcal{S}_{\bar C}$ denote the sets of capture states and non-capture states, respectively. For each state $u\in\mathcal{S}$, whether $u\in\mathcal{S}_C$ or not depends on the transmission probability of State-$u$ HOL batches. For illustration, we assume that the transmission probability after the $k$-th failure of a HOL batch is $q_k=1$ for $0\leq k \leq n_C-1$ and $q_k\ll 1$ for $n_C \leq k \leq K$, with which the number of capture states is equal to $n_C$. Specifically, for capture-free backoff, i.e., $n_C=0$, we have $\mathcal{S}_C=\emptyset$ and $\mathcal{S}_{\bar C}=\mathcal{S}=\{\text{T},\text{B}_0,\cdots,\text{B}_K\}$. For capture-based backoff, i.e., $n_C\geq 1$, we have $\mathcal{S}_C=\{\text{T},\text{B}_1,\cdots,\text{B}_{n_C-1}\}$ and $\mathcal{S}_{\bar C}=\{\text{B}_{n_C},\cdots,\text{B}_K\}$.\footnote{Here State B$_0$ is included in neither $\mathcal{S}_C$ nor $\mathcal{S}_{\bar C}$. Note that in the saturated condition with capture-based backoff, if there is a fresh HOL batch, no other HOL batches can be in a capture state. As a result, a fresh HOL batch always sees a non-reserved channel, i.e., $\beta^{(0)}=1$. With $q_0=1$, the probability of a HOL batch being at State B$_0$ is zero.}

For each HOL batch whose first packet is at the $k$-th backoff stage, the steady-state probability of the channel being not reserved, $\beta^{(k)}$, and the steady-state probability of successful transmission given that the channel is not reserved, $p^{(k)}$, can be written as
\begin{equation}
    \vspace{0.2cm}
    \beta^{(k)} =
    \begin{cases}
        \beta_C & 0\leq k \leq n_C\!-\!1 \\
        \beta_{\bar{C}} & n_C \leq k \leq K,
    \end{cases}
    \;p^{(k)} =
    \begin{cases}
        p_C & 0\leq k \leq n_C\!-\!1 \\
        p_{\bar{C}} & n_C \leq k \leq K,
    \end{cases}
    \label{eq3-2-2-1}
\end{equation}
where $\beta_C$ and $\beta_{\bar{C}}$ denote the steady-state probabilities that the channel is not reserved for each HOL batch in a capture state or a non-capture state, respectively, which can be obtained as
\begin{equation}
    \vspace{0.2cm}
    \beta_C = 1,\;\;\;\beta_{\bar{C}} = \tfrac{1}{1+\tfrac{(n-1)(M-1)p_{\bar{C}}\tilde{q}}{(1-p_C)^{n_C}}}.
    \label{eq3-2-2-2}
\end{equation}
$p_C$ and $p_{\bar{C}}$ denote the steady-state probabilities of successful transmission given that the channel is not reserved for each HOL batch in a capture state or a non-capture state, respectively, which can be obtained as
\begin{equation}
    \vspace{0.2cm}
    p_C = (1-\tilde{q})^{n-1},\;\;\;p_{\bar{C}} = \tfrac{p_C}{1+\tfrac{(n-1)\tilde{q}}{(1-p_C)^{n_C}}\bigl(1-(1-p_C)^{n_C}\bigr)}.
    \label{eq3-2-2-3}
\end{equation}
In (\ref{eq3-2-2-2})-(\ref{eq3-2-2-3}), $\tilde{q}$ is the average transmission probability for HOL batches in non-capture states:
\begin{equation}
    \vspace{0.2cm}
    \tilde{q}=\tfrac{1}{\frac{(1-p_{\bar{C}})^{K-n_C}}{q_K}+\sum_{k=n_C}^{K-1}\frac{p_{\bar{C}}(1-p_{\bar{C}})^{k-n_C}}{q_k}}.
    \label{eq3-2-2-4}
\end{equation}
The detailed derivation of (\ref{eq3-2-2-2})-(\ref{eq3-2-2-4}) is presented in Appendix \ref{sec_ap1}. Note that for connection-free Aloha with capture-based backoff, i.e., $M=1$ and $n_C\geq 1$, (\ref{eq3-2-2-2}) is reduced to $\beta_C=\beta_{\bar{C}}=1$, and (\ref{eq3-2-2-3}) is consistent with Eq. (20) and Eq. (22) in \cite{Aloha_capture_phenomenon}. For connection-based Aloha with capture-free backoff, i.e., $M>1$ and $n_C=0$, we have $\beta^{(k)}=\beta_{\bar{C}}=\tfrac{1}{1+(n-1)(M-1)p_{\bar{C}}\tilde{q}}$ and $p^{(k)}=p_{\bar{C}}=(1-\tilde{q})^{n-1}$ for $0\leq k \leq K$, which are consistent with Eq. (30) and Eq. (43) in \cite{10750858}, respectively.

Based on the preceding analysis, the network throughput and short-term fairness performance will be characterized in the following sections.

\subsection{Network Throughput}\label{sec4-3}

For each node, its successful transmissions occur if and only if its HOL batch is at State T. As a consequence, the throughput of each node is given by $\lambda_{out,i} = \tilde{\pi}_T$, $i\in\mathcal{N}$. The network throughput can then be obtained by combining (\ref{eq3-2-1-1})-(\ref{eq3-2-2-4}) with (\ref{eq1-3}) as\vspace{0.1cm}
\begin{equation}
    \hat{\lambda}_{out} = \tfrac{M}{M+\tfrac{1-p_C-(1-p_C)^{n_C}}{p_C}+\tfrac{(1-p_C)^{n_C}}{np_C\tilde{q}}},
    \label{eq4-3-2}
    \vspace{0.15cm}
\end{equation}
where $p_C$ and $\tilde{q}$ are given by (\ref{eq3-2-2-3}) and (\ref{eq3-2-2-4}), respectively.

It can be seen from (\ref{eq4-3-2}) that the network throughput $\hat{\lambda}_{out}$ is a monotonic increasing function of the batch size $M$, which indicates that the network throughput performance can be improved by transmitting more data packets after each success. For given $M$, (\ref{eq3-2-2-3})-(\ref{eq4-3-2}) suggest that the network throughput is closely determined by the transmission probabilities in non-capture states $q_{n_C},\cdots,q_K$, and can be maximized by optimally tuning them. Theorem \ref{theorem1} presents the maximum network throughput $\hat{\lambda}_{\max}=\sup_{\{q_{n_C},\cdots,q_K\}} \hat{\lambda}_{out}$.
\newtheorem{theorem}{\bf Theorem}
\begin{theorem}
    With the number of capture states $n_C=0$, the maximum network throughput $\hat{\lambda}_{\max}$ is given by\vspace{0.1cm}
    \begin{equation}
        \hat{\lambda}_{\max} = \tfrac{M}{M-1+\frac{1}{(1-\frac{1}{n})^{n-1}}},
        \label{eq4-3-3}
        \vspace{0.15cm}
    \end{equation}
    which is achieved when $\tilde{q}=\tfrac{1}{n}$.
    
    With $n_C\geq 1$, as $\tilde{q}\to 0$,\vspace{0.1cm}
    \begin{equation}
        \hat{\lambda}_{out} \to \hat{\lambda}_{\max} =
        \begin{cases}
            \tfrac{M}{M+\frac{n-1}{n}} & n_C=1 \\
            1 & n_C\geq 2.
        \end{cases}
        \label{eq4-3-5}
    \end{equation}
\label{theorem1}
\end{theorem}
\textit{Proof:} See Appendix \ref{sec_ap2}.

\begin{figure*}[!t]
    \normalsize
    \setcounter{mytempeqncnt}{\value{equation}}
    \setcounter{equation}{22}
    \begin{equation}
        \begin{split}
            G_{D}^{\prime\prime}(1) = & \sum_{j=0}^{K} \biggl(G_{Y_{B_j}}^{\prime\prime}(1)+ 2M G_{Y_{B_j}}^{\prime}(1) - 2 \bigl(G_{Y_{B_j}}^{\prime}(1)\bigr)^2 + 2 G_{Y_{B_j}}^{\prime}(1)\sum_{k=j}^{K} G_{Y_{B_{k}}}^{\prime}(1) \prod_{i=j}^{k-1}\bigl(1-p^{(i)}\bigr)\biggr)\cdot\prod_{i=0}^{j-1}\bigl(1-p^{(i)}\bigr) \\
            & -\beta^{(0)}q_0\biggl(G_{Y_{B_0}}^{\prime\prime}(1)+ 2M G_{Y_{B_0}}^{\prime}(1) - 2 \bigl(G_{Y_{B_0}}^{\prime}(1)\bigr)^2 + 2 G_{Y_{B_0}}^{\prime}(1)\sum_{k=0}^{K} G_{Y_{B_k}}^{\prime}(1) \prod_{i=0}^{k-1}\bigl(1-p^{(i)}\bigr)\biggr) + M(M-1).
        \end{split}
        \label{eq4-4-4}
    \end{equation}
    \setcounter{equation}{\value{mytempeqncnt}}
    \hrulefill
\end{figure*}

Theorem \ref{theorem1} reveals that the maximum network throughput $\hat{\lambda}_{\max}$ closely depends on connection-free or connection-based and capture-free or capture-based. Specifically, for connection-free Aloha (i.e., $M=1$), $\hat{\lambda}_{\max}$ is given by $(1-\tfrac{1}{n})^{n-1}$ with capture-free backoff. With capture-based backoff, $\hat{\lambda}_{\max}$ is equal to $\tfrac{n}{2n-1}$ for $n_C=1$, and 1 for $n_C\geq 2$. It suggests that for large $n$, the maximum network throughput of connection-free Aloha can be substantially increased from $e^{-1}$ to 0.5 by introducing one capture state, and further be increased to 1 by increasing the number of capture states to 2 or larger.

With connection-based Aloha, on the other hand, it can be seen from (\ref{eq4-3-3})-(\ref{eq4-3-5}) that as the batch size $M\to \infty$, the maximum network throughput $\hat{\lambda}_{\max}\to 1$ regardless of the number of capture states $n_C$. It implies that different from connection-free Aloha, introducing more capture states may not bring significant gains in the maximum network throughput for connection-based Aloha with a large $M$.

\subsection{Short-Term Fairness}\label{sec4-4}

Let $\bar{D}$ and $\sigma_{D}^2$ denote the mean and the variance of service time of HOL batches in each node's queue, respectively. The short-term fairness index $J_T$ defined in (\ref{eq1-4}) can be written as
\begin{equation}
    J_T \overset {n\gg 1, T\gg M}{\approx} \tfrac{1}{1+\tfrac{\sigma_{D}^2}{\bar{D}}\cdot\tfrac{1}{T}}.
    \label{eq4-4-1}
\end{equation}
The detailed derivation of (\ref{eq4-4-1}) is presented in Appendix \ref{sec_ap3}. Note that $\bar{D}$ and $\sigma_{D}^2$ are given by
\begin{equation}
    \bar{D} = G_{D}^{\prime}(1),\;\;\;\sigma_{D}^2 = G_{D}^{\prime\prime}(1) + G_{D}^{\prime}(1) - \bigl(G_{D}^{\prime}(1)\bigr)^2,
    \label{eq4-4-2}
\end{equation}
where $G_{D}(z)$ denotes the probability generating function of service time of HOL batches in each node's queue. Based on the Markov renewal process of HOL batches established in Section \ref{sec4-1}, $G_{D}^{\prime}(1)$ and $G_{D}^{\prime\prime}(1)$ can be obtained as
\begin{equation}
    G_{D}^{\prime}(1) = M - \beta^{(0)}q_0G_{Y_{B_0}}^{\prime}(1) + \sum_{j=0}^{K}G_{Y_{B_j}}^{\prime}(1)\prod_{i=0}^{j-1}(1-p^{(i)}),
    \label{eq4-4-3}
\end{equation}
and (\ref{eq4-4-4}), respectively, where $\beta^{(k)}$ and $p^{(k)}$ are given by (\ref{eq3-2-2-1})-(\ref{eq3-2-2-4}), and $Y_{B_k}$ denotes the sojourn time of a HOL batch at State B$_k$, which can be regarded as a geometrically distributed random variable with parameter $\beta^{(k)}q_k$ for $0 \leq k \leq K-1$, and $p^{(K)}\beta^{(K)}q_K$ for $k=K$. (\ref{eq4-4-4}) is shown at the top of this page. The detailed derivation of (\ref{eq4-4-3})-(\ref{eq4-4-4}) is shown in Appendix \ref{sec_ap4}. 

Recall that the maximum network throughput $\hat{\lambda}_{\max}$ given by (\ref{eq4-3-3}) and (\ref{eq4-3-5}) is independent of the transmission probabilities in non-capture states $q_{n_C},\cdots,$ $q_K$. The short-term fairness performance, however, is crucially determined by them. Intuitively, the smaller difference in transmission probabilities of nodes' HOL packets, the better fairness performance. Therefore, to optimize the short-term fairness performance, $q_k$ should be equal for $k\geq n_C$, i.e., $K=n_C$. With $K=n_C$, by substituting (\ref{eq3-2-2-1}) into (\ref{eq4-4-3})-(\ref{eq4-4-4}) and noting that $q_k=1$, $\beta^{(k)}=\beta_C=1$, $G_{Y_{B_k}}(z)=z$ for $0\leq k \leq n_C-1$ and $G_{Y_{B_{n_C}}}(z)=\tfrac{p_{\bar{C}}\beta_{\bar{C}}q_K \cdot z}{1-(1-p_{\bar{C}}\beta_{\bar{C}}q_K)z}$ for $k=n_C$, we have
\setcounter{equation}{23}
\begin{equation}
    G_{D}^{\prime}(1) = M + \tfrac{1-p_C}{p_C} + (1-p_C)^{n_C}\Bigl(\tfrac{1}{p_{\bar{C}}\beta_{\bar{C}}q_{n_C}}-\tfrac{1}{p_C}\Bigr),
    \label{eq4-4-5}
\end{equation}
and
\begin{equation}
    \begin{split}
        G_{D}^{\prime\prime}(1) = & M(M\!-\!1) \!+\! \tfrac{2(1-p_C)(M-1)}{p_C} \!+\! \tfrac{2(1-p_C)}{p_C^2} \!+\! 2(1\!-\!p_C)^{n_C}\cdot \\
        & \Bigl(\tfrac{1}{p_{\bar{C}}\beta_{\bar{C}}q_{n_C}}-\tfrac{1}{p_C}\Bigr)\Bigl(\tfrac{1}{p_{\bar{C}}\beta_{\bar{C}}q_{n_C}}+\tfrac{1}{p_C}+M+n_C-2\Bigr),
    \end{split}
    \label{eq4-4-6}
\end{equation}
where $\beta_{\bar{C}}$, $p_C$ and $p_{\bar{C}}$ can be derived by substituting $K=n_C$ into (\ref{eq3-2-2-2})-(\ref{eq3-2-2-4}). By combining (\ref{eq4-4-5})-(\ref{eq4-4-6}) with (\ref{eq4-4-2}), the mean $\bar{D}$ and the variance $\sigma_{D}^2$ of service time of HOL batches in each node's queue can be obtained, which are functions of the non-capture transmission probability $q_{n_C}$, the number of capture states $n_C$ and batch size $M$. The following theorem presents the effects of $q_{n_C}$, $n_C$ and $M$ on $\tfrac{\sigma_{D}^2}{\bar{D}}$.
\begin{theorem}
    With $K=n_C=0$,
    \begin{equation}
        \tfrac{\sigma_{D}^2}{\bar{D}} \!=\! \biggl(\!\tfrac{1}{q_0(1-q_0)^{n-1}}+(n-1)(M-1)\!\biggr)\biggl(\!1-\tfrac{M}{\tfrac{1}{q_0(1-q_0)^{n-1}}+n(M-1)}\!\biggr),
        \label{eq4-4-7}
    \end{equation}
    which is minimized as
    \begin{equation}
        \begin{split}
            \min_{0<q_0 \leq 1} \tfrac{\sigma_{D}^2}{\bar{D}} = &  \biggl(\tfrac{n}{(1-\frac{1}{n})^{n-1}}+(n-1)(M-1)\biggr)\biggl(1- \\
            & \tfrac{M}{\tfrac{n}{(1-\frac{1}{n})^{n-1}}+n(M-1)}\biggr)\overset{n\gg 1}{\approx} n(M+e-1),
        \end{split}    
    \label{eq4-4-8}
    \end{equation}
    when the transmission probability $q_0$ is set to $\tfrac{1}{n}$.
    
    With $K=n_C\geq 1$,
    \begin{equation}
        \tfrac{\sigma_{D}^2}{\bar{D}}=
        \begin{cases}
            \Theta\Bigl(\tfrac{1}{q_{n_C}^{n_C}}\Bigr) & \text{as } q_{n_C}\to 0 \\
            \Theta(M) & \text{as } M \to \infty.
        \end{cases}
        \label{eq4-4-10}
    \end{equation}
    \label{theorem2}
    \textit{Proof:} See Appendix \ref{sec_ap5}.
\end{theorem}
Note from (\ref{eq4-4-1}) that the short-term fairness index $J_T$ is a monotonic decreasing function of $\tfrac{\sigma_{D}^2}{\bar{D}}$. Theorem \ref{theorem2} reveals that the short-term fairness performance is critically determined by the transmission probability of non-capture state $q_{n_C}$, the number of capture states $n_C$ and batch size $M$.

\subsection{Tradeoff Between Network Throughput and Short-Term Fairness}\label{sec4-5}

The preceding analysis shows that the throughput and fairness performance closely depends on the access design features, such as connection-free or connection-based and capture-free or capture-based.

\subsubsection{Connection-Free Aloha ($M=1$)}\label{sec4-5-1}

It can be seen from Theorem \ref{theorem1} and Theorem \ref{theorem2} that for capture-free backoff with $K=n_C=0$, both the network throughput $\hat{\lambda}_{out}$ and short-term fairness index $J_T$ are maximized when the transmission probability $q_0=\tfrac{1}{n}$. In the capture-based case with $K=n_C\geq 1$, on the other hand, Theorem \ref{theorem1} shows that $\hat{\lambda}_{out}$ approaches the maximum $\hat{\lambda}_{\max}$ as the non-capture transmission probability $q_{n_C}\to 0$. Yet a decrease in $q_{n_C}$ leads to a lower $J_T$ for small $q_{n_C}$ because $\tfrac{\sigma_{D}^2}{\bar{D}}=\Theta\Bigl(\tfrac{1}{q_{n_C}^{n_C}}\Bigr)$, as Theorem \ref{theorem2} shows. This indicates that for connection-free Aloha with capture-based backoff, $q_{n_C}$ controls the tradeoff between the throughput and fairness performance.

\begin{figure}[t]
    \subfloat[]{
        \includegraphics[width=0.98\linewidth, height=0.36\linewidth]{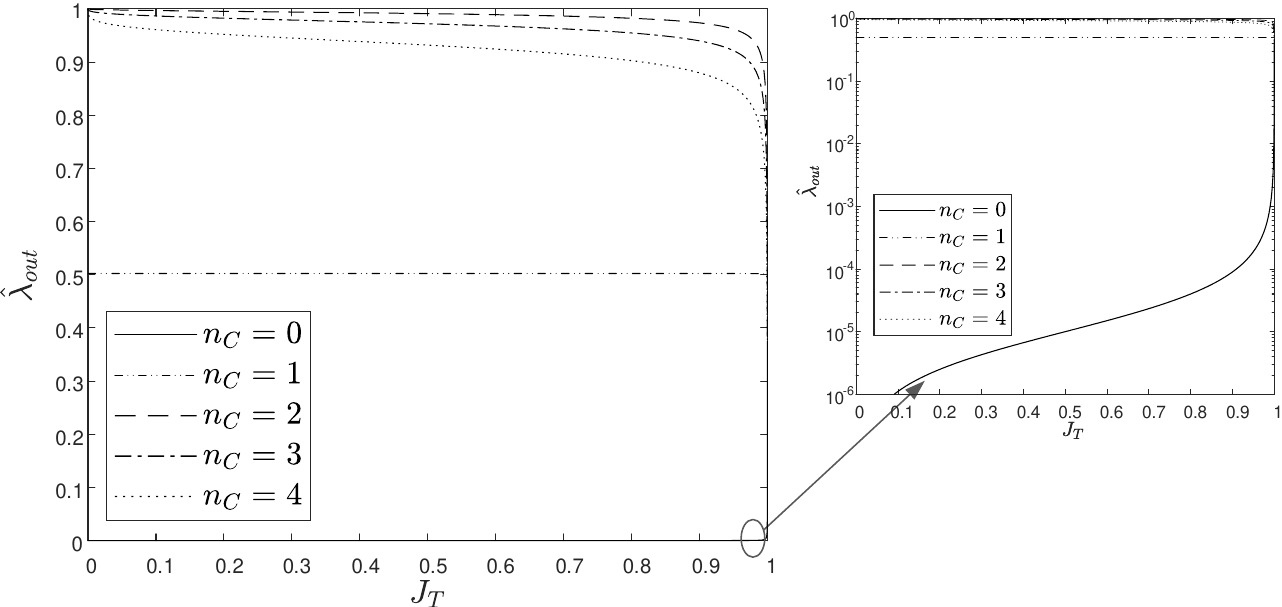}
        \label{fig4-5-1-1}
    }

    \subfloat[]{
        \includegraphics[width=0.98\linewidth, height=0.36\linewidth]{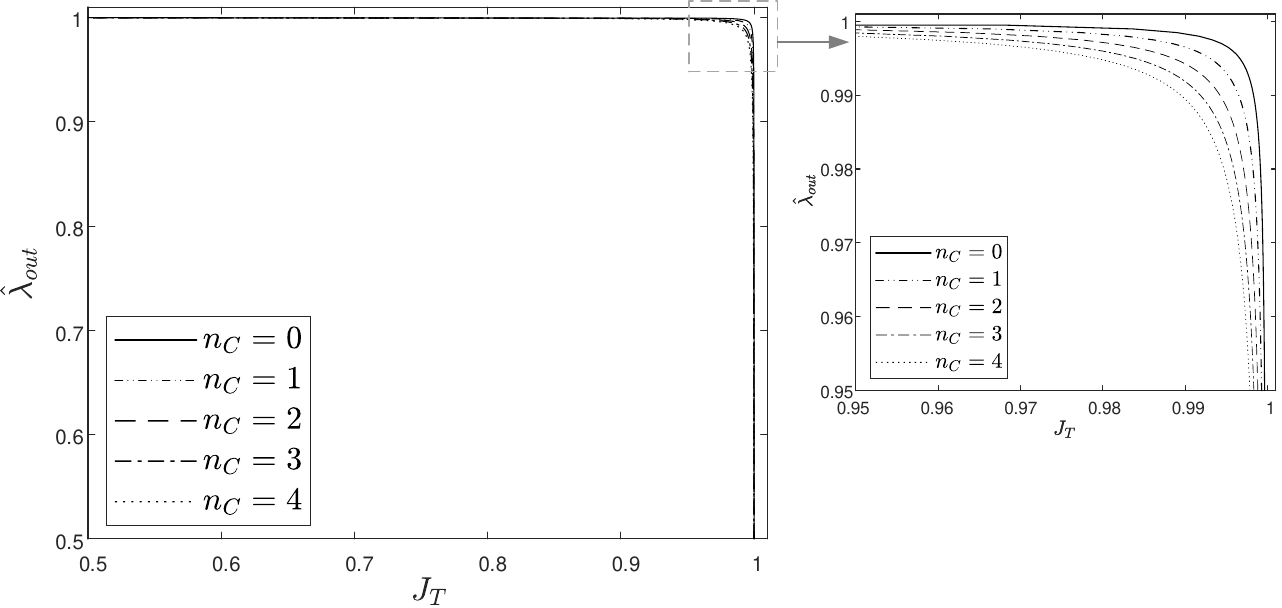} 
        \label{fig4-5-1-2}
    }
    \caption{Network throughput $\hat{\lambda}_{out}$ versus short-term fairness index $J_T$. $n=100$. $T=10^7$ slots. $K=n_C$. $n_C\in\{0,1,2,3,4\}$. (a) Connection-free Aloha. $q_{n_C}\in[10^{-10}, 10^{-2}]$. (b) Connection-based Aloha. $q_{n_C}\in[10^{-10}, 10^{-1}]$ and $M\in[1,10^6]$ are jointly optimized.}
    \label{fig4-5-1}
\end{figure}

Fig. \ref{fig4-5-1-1} illustrates how the network throughput $\hat{\lambda}_{out}$ varies with the short-term fairness index $J_T$ for connection-free Aloha, where $\hat{\lambda}_{out}$ and $J_T$ are obtained by varying the non-capture transmission probability $q_{n_C}$ from $10^{-10}$ to $10^{-2}$, and substituting $M=1$, $K=n_C$ into (\ref{eq3-2-2-2})-(\ref{eq4-3-2}), (\ref{eq4-4-1})-(\ref{eq4-4-2}), (\ref{eq4-4-5})-(\ref{eq4-4-6}). It can be seen that with $n_C=0$, the higher the short-term fairness index $J_T$, the larger the network throughput $\hat{\lambda}_{out}$. Both $\hat{\lambda}_{out}$ and $J_T$ are optimized when the transmission probability $q_0=\tfrac{1}{n}$.

For capture-based backoff with $K=n_C\geq 1$, on the other hand, we can clearly see the tradeoff between the throughput and short-term fairness performance in Fig. \ref{fig4-5-1-1}, which crucially depends on the number of capture states $n_C$. Recall from Theorems \ref{theorem1} and \ref{theorem2} that by increasing $n_C$ from 1 to 2, the maximum network throughput of connection-free Aloha can be substantially improved from 0.5 to 1 for large $n$. A further increase in $n_C$ brings no gain in the maximum network throughput $\hat{\lambda}_{\max}$, but leads to a larger $\tfrac{\sigma_{D}^2}{\bar{D}}$ and a smaller short-term fairness index $J_T$. Therefore, the optimal throughput-fairness tradeoff of connection-free Aloha is achieved when $n_C=2$.

\subsubsection{Connection-Based Aloha ($M>1$)}\label{sec4-5-2}

With connection-based Aloha, the network throughput $\hat{\lambda}_{out}$ and short-term fairness index $J_T$ are determined by both the non-capture transmission probability $q_{n_C}$ and batch size $M$. As demonstrated in Theorems \ref{theorem1} and \ref{theorem2}, for capture-free backoff with $K=n_C=0$, $\hat{\lambda}_{out}$ and $J_T$ are simultaneously optimized when the transmission probability is chosen as $q_0=\tfrac{1}{n}$, with which $M$ determines a tradeoff: an increase in $M$ leads to a larger maximum network throughput $\hat{\lambda}_{\max}$, but a lower short-term fairness index $J_T$ due to a larger $\tfrac{\sigma_D^2}{\bar{D}}$. For capture-based backoff with $K=n_C\geq 1$, on the other hand, we can see from Theorems \ref{theorem1} and \ref{theorem2} that with a small non-capture transmission probability $q_{n_C}$ or a large batch size $M$, a high network throughput can be achieved, while $\tfrac{\sigma_D^2}{\bar{D}}$ would be large, indicating poor short-term fairness performance. That is to say, $q_{n_C}$ and $M$ jointly control the throughput-fairness tradeoff of connection-based Aloha with capture-based backoff.

The throughput-fairness tradeoff for connection-based Aloha with different values of the number of capture states $n_C$ is illustrated Fig. \ref{fig4-5-1-2}. Specifically, in the case with $n_C=0$, we set the transmission probability $q_0$ to $\tfrac{1}{n}$ for optimizing the tradeoff. The corresponding network throughput $\hat{\lambda}_{out}$ and short-term fairness index $J_T$ by varying the batch size $M\in[1,10^6]$ can be obtained based on (\ref{eq4-3-3}), (\ref{eq4-4-1}), (\ref{eq4-4-8}). In the case with $n_C\geq 1$, by varying both the non-capture transmission probability $q_{n_C}\in[10^{-10},10^{-1}]$ and batch size $M\in[1,10^6]$, a set of throughput-fairness pairs $(\hat{\lambda}_{out}$, $J_T)$ can be derived from (\ref{eq3-2-2-2})-(\ref{eq4-3-2}), (\ref{eq4-4-1})-(\ref{eq4-4-2}), (\ref{eq4-4-5})-(\ref{eq4-4-6}), based on which the optimal throughput-fairness tradeoff can be obtained by finding the Pareto frontier of the set of $(\hat{\lambda}_{out}$, $J_T)$. Fig. \ref{fig4-5-1-2} shows that different from connection-free Aloha where the optimal tradeoff is achieved when $n_C=2$, with connection-based Aloha, the throughput-fairness tradeoff is the best when the number of capture states $n_C=0$, and becomes worse as $n_C$ grows. Intuitively, for connection-based Aloha with a large batch size $M$, introducing more capture states brings few gains in the throughput performance, yet worsens the fairness performance. That is why increasing $n_C$ cannot improve the throughput-fairness tradeoff.

\section{Throughput-Fairness Tradeoff for MTOA-L and MTOA-G}\label{sec5}

In this section, we will demonstrate how to apply the preceding queueing-theoretical analysis to find the optimal throughput-fairness tradeoff and the corresponding parameter setting for MTOA-L and MTOA-G, respectively.
\vspace{-0.2cm}
\subsection{MTOA-L}\label{sec5-1}

As shown in Section \ref{sec3}, with MTOA-L, the access strategy learned by nodes is indeed connection-free Aloha with the batch size $M=1$. For each HOL batch (packet), the transmission probability after its $k$-th failure is given by (\ref{eq3-1-1}) as $q_k=1$ for $0\leq k \leq n_C-1$ and $q_k=\tfrac{1}{L+1}$ for $k=K=n_C$, where $n_C$ is given by (\ref{eq3-1-2}) and dependent on the learning rate $\alpha$, the $Q$-value threshold $Q_{th}$, and the $Q$-value of transmission action seen by a fresh HOL packet $Q_0$. It can be seen from (\ref{eq3-1-2}) that with $Q_{th}\geq \alpha$ and $Q_{th}< \alpha$, the backoff strategy developed from MTOA-L is capture-free (i.e., $n_C=0$) and capture-based (i.e., $n_C\geq 1$), respectively. With $Q_{th}< \alpha$, we have $n_C=1$ for $\alpha=1$, and $n_C \approx \lceil \log_{1-\alpha}Q_{th} \rceil$ for $\alpha<1$ and small $1-\alpha$.\footnote{Specifically, given that a node has a fresh HOL packet at time slot $t$, it must make a successful transmission at the previous time slot. With $Q_{th}< \alpha$, its $Q$ value of the transmission action at time slot $t$ must be no smaller than $\alpha$ according to (\ref{eq2-1-6}). Thus, the $Q$-value seen by a fresh HOL packet $Q_0\geq \alpha$, and $Q_0\approx 1$ for small $1-\alpha$. By substituting $Q_0\approx 1$ into (\ref{eq3-1-2}), we then have $n_C\approx \lceil \log_{1-\alpha}Q_{th} \rceil$.}

\begin{figure*}[t]
    \centering
    \subfloat[]{
        \includegraphics[width=0.45\linewidth, height=0.25\linewidth]{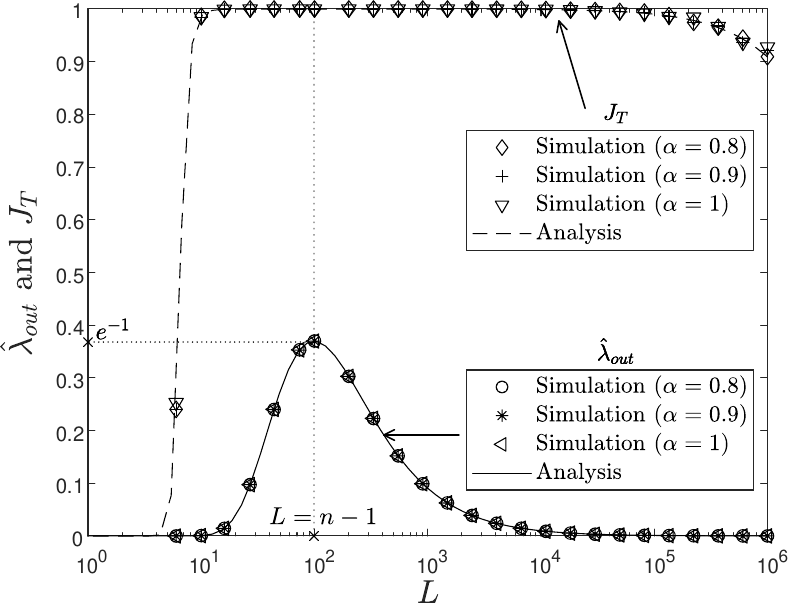}
        \label{fig5-1-1}
    }
    \subfloat[]{
        \includegraphics[width=0.45\linewidth, height=0.25\linewidth]{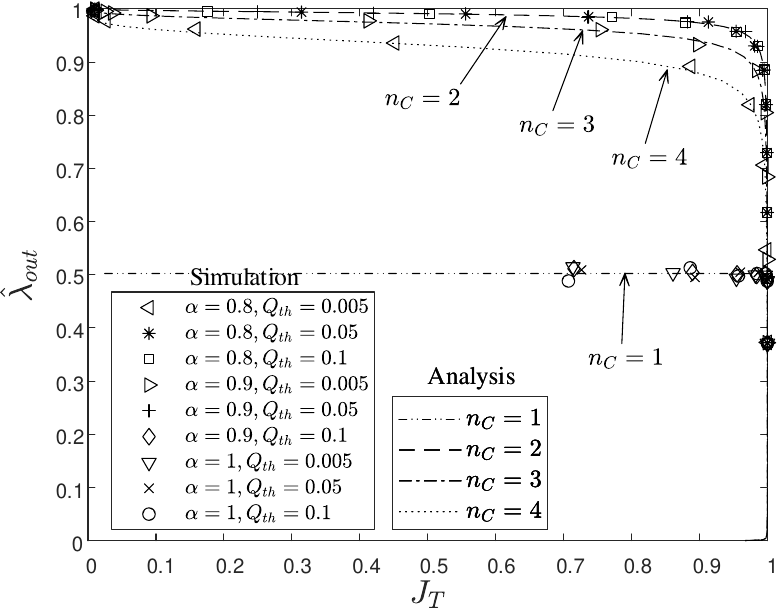}
        \label{fig5-1-2}
    }
    \caption{Analytical and simulated network throughput $\hat{\lambda}_{out}$ and short-term fairness index $J_T$ for MTOA-L. $n=100$. $T=10^7$ slots. $\alpha\in\{0.8,0.9,1\}$. (a) $\hat{\lambda}_{out}$ and $J_T$ versus the number of null actions $L$. $Q_{th}=1$. (b) $\hat{\lambda}_{out}$ versus $J_T$. $Q_{th}\in\{0.005,0.05,0.1\}$. $L\in[10^2,10^6]$.}
    \label{fig5-1}
    \vspace{-0.2cm}
\end{figure*}

It has been shown in Section \ref{sec4-3} and \ref{sec4-4} that for connection-free Aloha, the network throughput $\hat{\lambda}_{out}$ and short-term fairness index $J_T$ are crucially determined by the non-capture transmission probability $q_{n_C}$. Recall that with $K=n_C=0$, $\hat{\lambda}_{out}$ and $J_T$ are simultaneously maximized when $q_0=\tfrac{1}{n}$. It suggests that for MTOA-L with $Q_{th}\geq \alpha$, the number of null actions $L$ should be set to $n-1$ for optimizing the throughput and fairness performance. To verify that, both the analytical and simulated results of network throughput $\hat{\lambda}_{out}$ and short-term fairness index $J_T$ for MTOA-L are presented in Fig. \ref{fig5-1-1}, where the simulation is conducted under the setting $Q_{th}\geq \alpha$, and the analytical results of $\hat{\lambda}_{out}$ and $J_T$ are obtained by substituting $M=1$, $K=n_C=0$, $q_0=\tfrac{1}{L+1}$ into (\ref{eq3-2-2-3})-(\ref{eq4-3-2}) and (\ref{eq4-4-1}), (\ref{eq4-4-7}), respectively. It can be observed that the analysis well agrees with the simulation results. We can also see that when the number of null actions $L=n-1$, $\hat{\lambda}_{out}$ is maximized as $e^{-1}$, while $J_T$ is close to 1, indicating good fairness performance.

With $K=n_C\geq 1$, on the other hand, as the non-capture transmission probability $q_{n_C}\to 0$, $\hat{\lambda}_{out}$ approaches the maximum, while $J_T$ decreases. It means that for MTOA-L with $Q_{th}< \alpha$, the number of null actions $L$ controls the tradeoff between the throughput and fairness performance. Section \ref{sec4-5} has shown that the tradeoff with connection-free Aloha is closely dependent on the number of capture states $n_C$, and optimized when $n_C=2$. This indicates that to optimize the throughput-fairness tradeoff of MTOA-L, the learning rate $\alpha$ and $Q$-value threshold $Q_{th}$ need to be properly chosen, i.e., $Q_{th}$ and $\alpha$ should satisfy $Q_{th}<\alpha<1$ and $\lceil \log_{1-\alpha}Q_{th} \rceil=2$ for $\alpha$ close to 1.

Fig. \ref{fig5-1-2} illustrates the simulated and analytical network throughput $\hat{\lambda}_{out}$ and short-term fairness index $J_T$ for MTOA-L under different settings of $\alpha$ and $Q_{th}$, where the simulated throughput-fairness pairs are obtained by varying the number of null actions $L$ from $10^2$ to $10^6$, and the analytical tradeoff curves are obtained by combining $M=1$, $K=n_C$ with (\ref{eq3-2-2-2})-(\ref{eq4-3-2}), (\ref{eq4-4-1})-(\ref{eq4-4-2}), (\ref{eq4-4-5})-(\ref{eq4-4-6}). We can see that the simulation results with different values of $\alpha$ and $Q_{th}$ fall onto the corresponding analytical curves. The results in Fig. \ref{fig5-1} also corroborate that the optimal throughput-fairness tradeoff of MTOA-L is achieved by properly choosing the learning rate $\alpha$ and $Q$-value threshold $Q_{th}$ to satisfy $Q_{th}<\alpha<1$ and $\lceil \log_{1-\alpha}Q_{th} \rceil=2$.
\vspace{-0.2cm}
\subsection{MTOA-G}\label{sec5-2}

Recall that with MTOA-G, the learned access strategy is shown in Section \ref{sec3} as connection-based Aloha with the batch size $M>1$. For each batch, the first data packet is transmitted with a constant probability $\tfrac{1}{L+1}$, which is a capture-free backoff strategy with $K=n_C=0$ and $q_0=\tfrac{1}{L+1}$.

For connection-based Aloha with $K=n_C=0$, Theorem \ref{theorem1} and Theorem \ref{theorem2} have shown that both the network throughput $\hat{\lambda}_{out}$ and short-term fairness index $J_T$ are maximized when the transmission probability $q_0=\tfrac{1}{n}$. It suggests that with MTOA-G, the number of null actions $L$ should be set to $n-1$ to optimize both the throughput and fairness performance. Further note from Theorems \ref{theorem1} and \ref{theorem2} that for connection-based Aloha with capture-free backoff, when the transmission probability $q_0$ is optimally selected, an increase in the batch size $M$ leads to a higher maximum network throughput $\hat{\lambda}_{\max}$ but a lower short-term fairness index $J_T$. It indicates that with MTOA-G, $M$ determines the optimal tradeoff between the throughput and fairness performance.

\begin{figure*}[t]
    \centering
    \subfloat[]{
        \includegraphics[width=0.45\linewidth, height=0.25\linewidth]{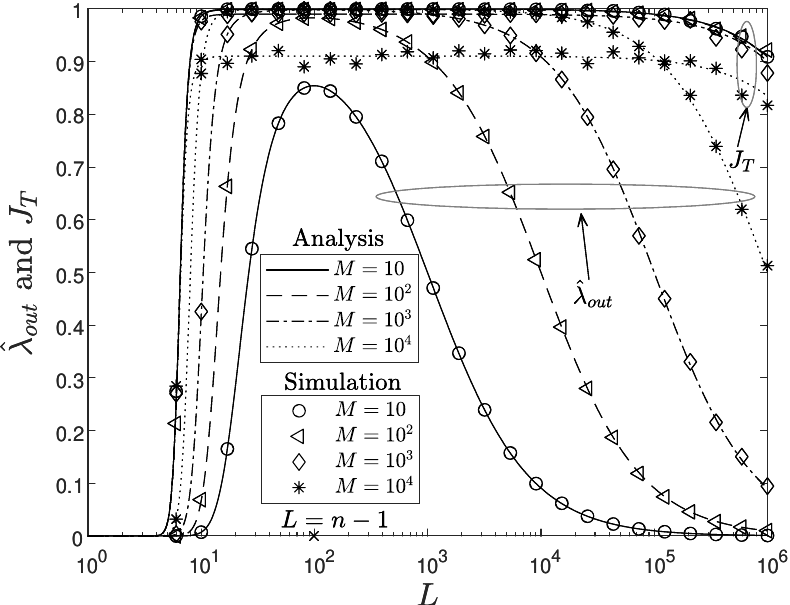}
        \label{fig5-2-1}
    }
    \subfloat[]{
        \includegraphics[width=0.45\linewidth, height=0.25\linewidth]{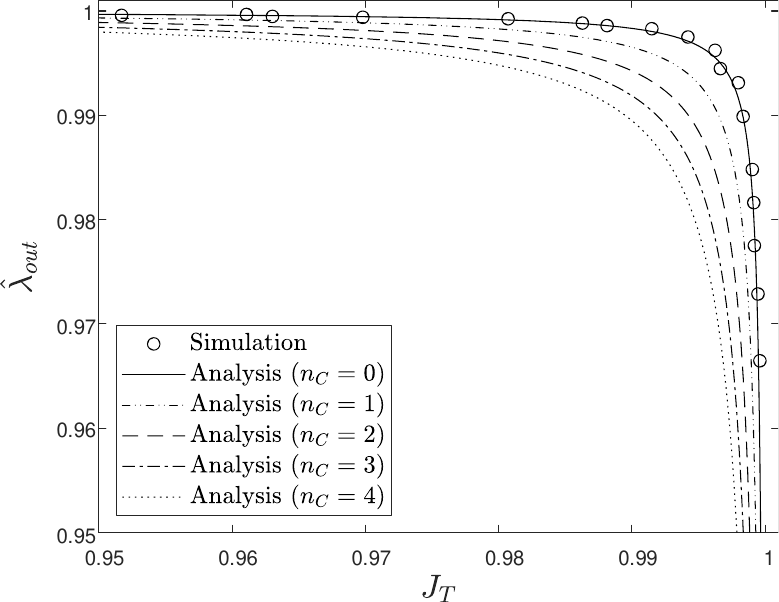}
        \label{fig5-2-2}
    }
    \caption{Analytical and simulated network throughput $\hat{\lambda}_{out}$ and short-term fairness index $J_T$ for MTOA-G. $n=100$. $T=10^7$ slots. $\alpha=0.9$. (a) $\hat{\lambda}_{out}$ and $J_T$ versus the number of null actions $L$. $M\in\{10,10^2,10^3,10^4\}$. (b) $\hat{\lambda}_{out}$ versus $J_T$. $L=n-1=99$. $M\in[50,5000]$.}
    \label{fig5-2}
    \vspace{-0.1cm}
\end{figure*}

Fig. \ref{fig5-2-1} illustrates how the network throughput $\hat{\lambda}_{out}$ and short-term fairness index $J_T$ of MTOA-G vary with the number of null actions $L$, where the analytical results of $\hat{\lambda}_{out}$ and $J_T$ are obtained by substituting $K=n_C=0$, $q_0=\tfrac{1}{L+1}$ into (\ref{eq3-2-2-3})-(\ref{eq4-3-2}) and (\ref{eq4-4-1}), (\ref{eq4-4-7}), respectively. It can be observed that both $\hat{\lambda}_{out}$ and $J_T$ are maximized when the number of null actions $L=n-1$. With $L=n-1$, an increase in the batch size $M$ leads to improved throughput performance, but worse short-term fairness.

Fig. \ref{fig5-2-2} further presents the simulated throughput-fairness pairs of MTOA-G with the number of null actions $L=n-1$ and the batch size $M$ ranging from 50 to 5000. For illustration, the analytical tradeoff curves of connection-based Aloha are also plotted with different values of the number of capture states $n_C$. We can see that the simulation results fall onto the analytical curve with $n_C=0$. Recall that the throughput-fairness tradeoff of connection-based Aloha is optimal when $n_C=0$, which suggests that the capture-free backoff strategy learned via MTOA-G indeed achieves the optimal throughput-fairness tradeoff of connection-based Aloha.
\vspace{-0.3cm}
\subsection{Comparison of Optimal Throughput-Fairness Tradeoff with MTOA-L and MTOA-G}\label{sec5-3}

It has been demonstrated that to achieve the optimal tradeoff for MTOA-L, the learning rate $\alpha$ and $Q$-value threshold $Q_{th}$ should be carefully selected such that the number of capture states $n_C=2$. For MTOA-G, the tradeoff is optimized when the number of null actions $L$ is set to $n-1$.

\begin{figure*}[t]
    \centering
    \includegraphics[width=0.9\linewidth, height=0.27\linewidth]{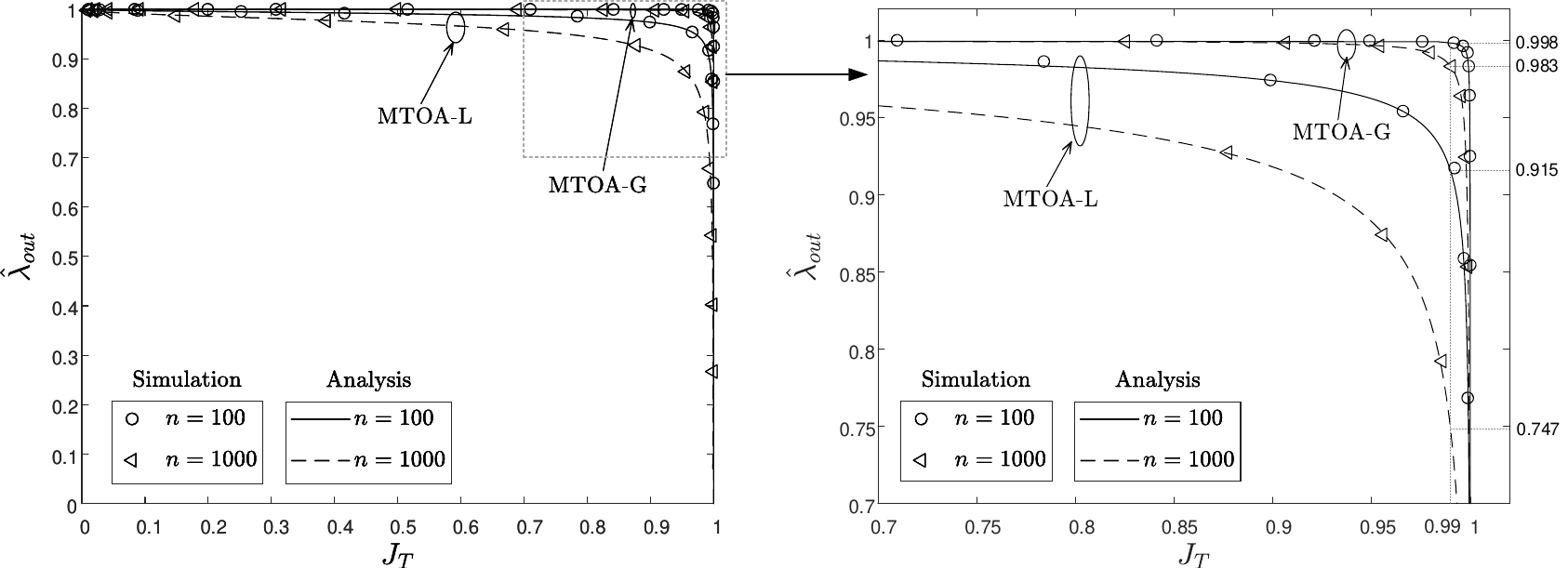}
    \caption{Analytical and simulated network throughput $\hat{\lambda}_{out}$ and short-term fairness index $J_T$ for MTOA-L with $Q_{th}=0.05$ and $L\in[10,10^6]$ and for MTOA-G with $L=n-1$ and $M\in[10,10^6]$. $n\in\{100,1000\}$. $T=10^7$ slots. $\alpha=0.9$.}
    \label{fig5-2-3}
\end{figure*}

In Fig. \ref{fig5-2-3}, the optimal throughput-fairness tradeoffs with MTOA-L or MTOA-G are compared, where both analytical and simulated results are presented. It can be seen that MTOA-G outperforms MTOA-L in terms of the optimal tradeoff thanks to the adoption of global rewards. Note that a better throughput-fairness tradeoff indicates a larger maximum network throughput that can be achieved for a given fairness constraint. As we can see from Fig. \ref{fig5-2-3}, if the short-term fairness index $J_T$ is required to be no smaller than 0.99 for $T=10^7$, then the maximum achievable network throughput with MTOA-G and MTOA-L can be found as 0.998 and 0.915, respectively, for the number of nodes $n=100$. As $n$ grows from $100$ to $1000$, the maximum achievable network throughput with MTOA-G is slightly reduced to 0.983, while that with MTOA-L sharply drops to 0.747. It suggests that the improvement in throughput-fairness tradeoff brought by MTOA-G would be especially significant when the number of nodes is large.
\vspace{-0.1cm}
\section{Conclusion}\label{sec6}

In this paper, we propose two MAB-based random access schemes for throughput maximization, and leverage the queueing-theoretical analysis to optimize their throughput-fairness tradeoffs. Specifically, based on the MAB framework, MTOA-L with local rewards and MTOA-G with global rewards are developed, and both shown to be able to achieve the maximum network throughput of 1, though with different short-term fairness performance. To further analyze and optimize the throughput-fairness tradeoff, the access strategies learned via MTOA-L and MTOA-G are identified, with key features extracted as connection-free or connection-based Aloha and capture-free or capture-based backoff. By establishing a unified queueing-theoretical framework to incorporate the above design features, the network throughput and short-term fairness of MTOA-L and MTOA-G can be characterized, and the setting of key parameters for optimizing the throughput-fairness tradeoff can be obtained for both MTOA-L and MTOA-G.

The analysis shows that for MTOA-L, the number of null actions $L$ determines the tradeoff between the network throughput and short-term fairness, to optimize which the learning rate $\alpha$ and $Q$-value threshold $Q_{th}$ should be properly chosen. For MTOA-G, on the other hand, the tradeoff is controlled by the batch size $M$ and optimized when the number of null actions $L$ is set to $n-1$. The comparison of the optimal throughput-fairness tradeoffs with MTOA-L and MTOA-G further suggests that though both are throughput-optimal, a much better throughput-fairness tradeoff can be achieved by MTOA-G, and the gain is enlarged as the number of nodes in the network grows.

\appendices

\section{Proof of Theorem \ref{theorem1}}\label{sec_ap2}

\begin{IEEEproof}
Let $g(\tilde{q})=\tfrac{1-(1-p_C)^{n_C}}{p_C}$ and $h(\tilde{q})=\tfrac{(1-p_C)^{n_C}}{p_C \tilde{q}}$. The network throughput $\hat{\lambda}_{out}$ can be rewritten from (\ref{eq4-3-2}) as
\begin{equation}
    \hat{\lambda}_{out} = \tfrac{M}{M-1+g(\tilde{q})+\tfrac{1}{n}h(\tilde{q})},
    \label{eq-ap2-2}
\end{equation}
where $\tilde{q}=\tfrac{1}{\frac{(1-p_{\bar{C}})^{K-n_C}}{q_K}+\sum_{k=n_C}^{K-1}\frac{p_{\bar{C}}(1-p_{\bar{C}})^{k-n_C}}{q_k}}$.

For $n_C=0$, we have $g(\tilde{q})=0$ and $h(\tilde{q})=\tfrac{1}{\tilde{q}(1-\tilde{q})^{n-1}}$ according to (\ref{eq3-2-2-3}). By noting that $h^{\prime}(\tilde{q})=\tfrac{n\tilde{q}-1}{\tilde{q}^2(1-\tilde{q})^{n}}$, we can obtain that $h(\tilde{q})$ for $\tilde{q}\in(0,1]$ is minimized as $\tfrac{n}{(1-\frac{1}{n})^{n-1}}$ when $\tilde{q}=\tfrac{1}{n}$. Also note from (\ref{eq-ap2-2}) that the network throughput $\hat{\lambda}_{out}$ monotonically decreases with $h(\tilde{q})$. As a consequence, with $n_C=0$, the network throughput $\hat{\lambda}_{out}$ is maximized as $\tfrac{M}{M-1+\tfrac{1}{(1-\frac{1}{n})^{n-1}}}$ when $\tilde{q}=\tfrac{1}{n}$.

For $n_C\geq 1$, on the other hand, note that $g(\tilde{q})=\sum_{k=0}^{n_C-1}(1-p_C)^k$, where $p_C$ is given by (\ref{eq3-2-2-3}) as $p_C=(1-\tilde{q})^{n-1}$, which is a monotonic decreasing function of $\tilde{q}$. Thus, $g(\tilde{q})$ is a monotonic non-decreasing function of $\tilde{q}$. For $h(\tilde{q})$, its first-order derivative with respect to $\tilde{q}$ is given by
\begin{equation}
    h^{\prime}(\tilde{q}) = -\tfrac{(1-p_C)^{n_C-1}}{\tilde{q}^2p_C^2}((n_C-1)\tilde{q}p_C p_C^{\prime} + p_C(1-p_C) + \tilde{q}p_C^{\prime}),
    \label{eq-ap2-3}
\end{equation}
where $p_C^{\prime}$ is the first-order derivation of $p_C$ with respect to $\tilde{q}$, and given by $p_C^{\prime}=-\tfrac{n-1}{1-\tilde{q}}p_C$ according to (\ref{eq3-2-2-3}). As $p_C(1-p_C) + \tilde{q}p_C^{\prime}<p_C\cdot\bigl((n-1)\tilde{q}-\tfrac{(n-1)\tilde{q}}{1-\tilde{q}}\bigr)<0$, we have $h^{\prime}(\tilde{q})>0$ for $n_C\geq 1$ according to (\ref{eq-ap2-3}). By noting from (\ref{eq-ap2-2}) that $\hat{\lambda}_{out}$ monotonically decreases with $g(\tilde{q})$ and $h(\tilde{q})$, we can conclude that with $n_C\geq 1$, the network throughput $\hat{\lambda}_{out}$ is a monotonic decreasing function of $\tilde{q}$.

Further note from (\ref{eq3-2-2-3}) that as $\tilde{q}\to 0$, $p_C\to 1$ and
\begin{equation}
    \tfrac{(1-p_C)^{n_C}}{(n-1)\tilde{q}} \to
    \begin{cases}
        1 & n_C = 1 \\
        0 & n_C \geq 2.
    \end{cases}
    \label{eq-ap2-4}
\end{equation}
By combining (\ref{eq-ap2-2}) with (\ref{eq-ap2-4}), we can then obtain that the maximum network throughput $\hat{\lambda}_{\max}=\tfrac{M}{M+\frac{n-1}{n}}$ for $n_C=1$ and $\hat{\lambda}_{\max}=1$ for $n_C\geq 2$.
\end{IEEEproof}
\vspace{-0.2cm}
\section{Proof of Theorem \ref{theorem2}}\label{sec_ap5}

\begin{IEEEproof}
To prove Theorem \ref{theorem2}, let us start from the case with $K=n_C=0$.

\subsubsection{$K=n_C=0$}\label{sec_ap5_1}

By substituting $K=n_C=0$ into (\ref{eq3-2-2-2})-(\ref{eq3-2-2-4}), (\ref{eq4-4-2}), (\ref{eq4-4-5})-(\ref{eq4-4-6}), $\tfrac{\sigma_{D}^2}{\bar{D}}$ can be obtained as (\ref{eq4-4-7}). Let $f(q_0) = q_0(1-q_0)^{n-1}$. $\tfrac{\sigma_{D}^2}{\bar{D}}$ can be rewritten from (\ref{eq4-4-7}) as
\begin{equation}
    \tfrac{\sigma_{D}^2}{\bar{D}} = \biggl(\tfrac{1}{f(q_0)}+(n-1)(M-1)\biggr)\biggl(1-\tfrac{M}{\tfrac{1}{f(q_0)}+n(M-1)}\biggr).
    \label{eq-ap5-1-1}
\end{equation}
It can be seen from (\ref{eq-ap5-1-1}) that $\tfrac{\sigma_{D}^2}{\bar{D}}$ monotonically decreases with $f(q_0)$. By noting that $f^{\prime}(q_0) = (1-q_0)^{n-2}(1-nq_0)$, we can obtain that for $0 < q_0 \leq 1$, when $q_0=\tfrac{1}{n}$, $f(q_0)$ is maximized, and thus $\tfrac{\sigma_{D}^2}{\bar{D}}$ is minimized. By substituting $q_0=\tfrac{1}{n}$ into (\ref{eq-ap5-1-1}), $\min_{0<q_0\leq 1} \tfrac{\sigma_{D}^2}{\bar{D}}$ can be obtained as (\ref{eq4-4-8}), which can further be approximated as $n(M+e-1)$ by applying $n-1\approx n$ and $(1-x)^n\approx e^{-nx}$ for $0<x<1$ when $n\gg 1$.

\subsubsection{$K=n_C\geq 1$}\label{sec_ap5_2}

For $\tfrac{\sigma_{D}^2}{\bar{D}}$ with $K=n_C\geq 1$, let us first determine the scaling orders of $G_D^{\prime}(1)$ and $G_D^{\prime\prime}(1)$.

Note from (\ref{eq3-2-2-3})-(\ref{eq3-2-2-4}) that with $K=n_C$, $p_C\to 1$ as $q_{n_C}\to 0$. For given $M$, the scaling orders of $G_D^{\prime}(1)$ and $G_D^{\prime\prime}(1)$ in terms of $q_{n_C}$ are therefore determined by $\tfrac{(1-p_C)^{n_C}}{p_{\bar{C}}\beta_{\bar{C}}q_{n_C}}$ according to (\ref{eq4-4-5}), and by $\tfrac{2(1-p_C)^{n_C}}{(p_{\bar{C}}\beta_{\bar{C}}q_{n_C})^2}$ and $\tfrac{2(M+n_C-2)(1-p_C)^{n_C}}{p_{\bar{C}}\beta_{\bar{C}}q_{n_C}}$ according to (\ref{eq4-4-6}), respectively. By noting that $\tfrac{1}{p_{\bar{C}}\beta_{\bar{C}}q_{n_C}}=\tfrac{n-1}{(1-p_C)^{n_C}}\Bigl(M+\tfrac{1-p_C}{p_C}-\tfrac{(1-p_C)^{n_C}}{p_C}\Bigr)+\tfrac{1}{p_C q_{n_C}}$ according to (\ref{eq3-2-2-2})-(\ref{eq3-2-2-3}) and $p_C\to 1$ as $q_{n_C}\to 0$, we have
\begin{equation}
    \lim_{q_{n_C}\to 0} \tfrac{(1-p_C)^{n_C}}{p_{\bar{C}}\beta_{\bar{C}}q_{n_C}} = (n-1)M + \lim_{q_{n_C}\to 0} \tfrac{(1-p_C)^{n_C}}{q_{n_C}},
    \label{eq-ap5-1}
\end{equation}
and
\begin{equation}
    \lim_{q_{n_C}\to 0} \tfrac{(1-p_C)^{n_C}}{(p_{\bar{C}}\beta_{\bar{C}}q_{n_C})^2} = \tfrac{2(n-1)M}{q_{n_C}} + \tfrac{(n-1)^2M^2}{(1-p_C)^{n_C}} + \lim_{q_{n_C}\to 0} \tfrac{(1-p_C)^{n_C}}{q_{n_C}^2}.
    \label{eq-ap5-2}
\end{equation}
Further note that as $q_{n_C}\to 0$, $(1-p_C)^{n_C} = \Theta (q_{n_C}^{n_C})$. By combining (\ref{eq-ap5-1})-(\ref{eq-ap5-2}), we have
\begin{equation}
    \tfrac{(1-p_C)^{n_C}}{p_{\bar{C}}\beta_{\bar{C}}q_{n_C}} = \Theta (1),\;\;\;\tfrac{(1-p_C)^{n_C}}{(p_{\bar{C}}\beta_{\bar{C}}q_{n_C})^2} = \Theta \Big(\tfrac{1}{q_{n_C}^{n_C}}\Big),
    \label{eq-ap5-3}
\end{equation}
as $q_{n_C}\to 0$. The scaling orders of $G_D^{\prime}(1)$ and $G_D^{\prime\prime}(1)$ in terms of $q_{n_C}$ can then be obtained from (\ref{eq-ap5-3}) as $\Theta (1)$ and $\Theta \Big(\tfrac{1}{q_{n_C}^{n_C}}\Big)$, respectively. According to (\ref{eq4-4-2}), we have $\bar{D}=\Theta (1)$ and $\sigma_{D}^2=\Theta \Big(\tfrac{1}{q_{n_C}^{n_C}}\Big)$, and thus $\tfrac{\sigma_{D}^2}{\bar{D}}=\Theta\Bigl(\tfrac{1}{q_{n_C}^{n_C}}\Bigr)$ as $q_{n_C}\to 0$.

For the scaling orders of $G_D^{\prime}(1)$ and $G_D^{\prime\prime}(1)$ in terms of the batch size $M$, on the other hand, we can obtain from (\ref{eq4-4-5})-(\ref{eq4-4-6}) that they are determined by $nM$ and $\tfrac{2(n-1)^2 M^2}{(1-p_C)^{n_C}}+2(n-1)M^2+M^2$, respectively, by noting that $\tfrac{1}{p_{\bar{C}}\beta_{\bar{C}}q_{n_C}}=\tfrac{n-1}{(1-p_C)^{n_C}}\Bigl(M+\tfrac{1-p_C}{p_C}-\tfrac{(1-p_C)^{n_C}}{p_C}\Bigr)+\tfrac{1}{p_C q_{n_C}}$. We can then obtain that as $M\to\infty$, the scaling orders of $G_D^{\prime}(1)$ and $G_D^{\prime\prime}(1)$ are $\Theta (M)$ and $\Theta(M^2)$, respectively. According to (\ref{eq4-4-2}), we have $\bar{D}=\Theta (M)$ and $\sigma_{D}^2=\Theta(M^2)$, and thus $\tfrac{\sigma_{D}^2}{\bar{D}}=\Theta (M)$ as $M\to\infty$.
\end{IEEEproof}

\section{Derivation of (\ref{eq3-2-2-2})-(\ref{eq3-2-2-4})}\label{sec_ap1}

In this appendix, we will derive the steady-state probability that the channel is not reserved $\beta^{(k)}$ and the steady-state probability of successful transmission conditioned on the channel not being reserved $p^{(k)}$ for each HOL batch whose first packet is at the $k$-th backoff stage.

\subsubsection{Derivation of $\beta^{(k)}$}

For each node, define event $E_1^{(k)}$ as the event that its HOL batch is in the backoff stage $k$, and $E_2$ as the event that the channel is reserved by one of the other $n-1$ nodes, $0 \leq k \leq K$. The steady-state probability that the channel is not reserved for each HOL batch whose first packet is at the backoff stage $k$ can then be written as
\begin{equation}
    \beta^{(k)} = 1- \Pr\{E_2 | E_1^{(k)}\} = 1 - \tfrac{\Pr\{E_2\}\Pr\{E_1^{(k)} | E_2\}}{\Pr\{E_1^{(k)}\}}.
    \label{eq-ap1-1}
\end{equation}

Note that the channel is reserved if and only if one node's HOL batch is in the last $M-1$ time slots of State T. If a node's HOL batch is in a capture state, no other nodes could have State-T HOL batches. Therefore, for $0 \leq k \leq n_C-1$, we have $\Pr\{E_2 | E_1^{(k)}\}=0$, and thus $\beta^{(k)}=\beta_C=1$.

Let us derive $\beta^{(k)}$ for $n_C \leq k \leq K$. The probability that the channel is reserved by one of the other $n-1$ nodes $\Pr\{E_2\}$ can be written as
\begin{equation}
    \Pr\{E_2\} = \tfrac{(n-1)(M-1)\tilde{\pi}_T}{M},
    \label{eq-ap1-2}
\end{equation}
by noting that the channel can be reserved by at most one node at each time slot. For a node, given that the channel is reserved by another node, its HOL batch must be at one of the following states: 1) State B$_0$, $\cdots$, B$_K$, and the first time slot of State T with capture-free backoff, or 2) State B$_{n_C}$, $\cdots$, B$_K$ with capture-based backoff. $\Pr\{E_1^{(k)} | E_2\}$ for $n_C \leq k \leq K$ can then be written as
\begin{equation}
    \begin{split}
        & \Pr\{E_1^{(k)} | E_2\} =
        \begin{cases}
            \tfrac{\frac{\tilde{\pi}_T}{M}+\tilde{\pi}_{B_0}}{\frac{\tilde{\pi}_T}{M}+\sum_{l=0}^{K}\tilde{\pi}_{B_l}} & k = 0 \\
            \tfrac{\tilde{\pi}_{B_k}}{\frac{\tilde{\pi}_T}{M}+\sum_{l=0}^{K}\tilde{\pi}_{B_l}} & 1 \leq k \leq K,
        \end{cases} \\
        & \Pr\{E_1^{(k)} | E_2\} = \tfrac{\tilde{\pi}_{B_k}}{\sum_{l=n_C}^{K}\tilde{\pi}_{B_l}},   
    \end{split}
    \label{eq-ap1-3}
\end{equation}
\vspace{-0.1cm}for capture-free and capture-based backoff, respectively. The probability that a node's HOL batch is at the backoff stage $k$, $n_C \leq k \leq K$, is given by
\begin{equation}
    \Pr\{E_1^{(k)}\} =
    \begin{cases}
        \frac{\tilde{\pi}_T}{M}+\tilde{\pi}_{B_0} & k = 0 \\
        \tilde{\pi}_{B_k} & 1 \leq k \leq K,
    \end{cases}
    \;\;\;\Pr\{E_1^{(k)}\}=\tilde{\pi}_{B_k},
    \label{eq-ap1-4}
\end{equation}
for capture-free backoff and capture-based backoff, respectively. By noting from (\ref{eq3-2-1-1})-(\ref{eq3-2-1-3}) that $\tfrac{\tilde{\pi}_T}{M}+\tilde{\pi}_{B_0}=\tfrac{\tilde{\pi}_T}{\beta^{(0)}q_0M}$, $\tilde{\pi}_{B_l}=\tfrac{\prod_{m=0}^{l-1}(1-p^{(m)})}{\beta^{(l)}q_l}\cdot\tfrac{\tilde{\pi}_T}{M}$ for $1\leq l \leq K-1$, and $\tilde{\pi}_{B_K}=\tfrac{\prod_{m=0}^{K-1}(1-p^{(m)})}{p^{(K)}\beta^{(K)}q_K}\cdot\tfrac{\tilde{\pi}_T}{M}$ and combining with (\ref{eq-ap1-1})-(\ref{eq-ap1-4}), $\beta^{(k)}$ for $n_C \leq k \leq K$ can be written as $\beta^{(k)} = \beta_{\bar{C}}$, where
\begin{equation}
        \beta_{\bar{C}} = 1 \!-\! \tfrac{(n-1)(M-1)}{\sum_{l=n_C}^{K-1}\!\tfrac{1}{\beta^{(\!l\!)}q_l}\!\prod_{m=0}^{l-1}(1-p^{(\!m\!)}) + \tfrac{1}{p^{(\!K\!)}\beta^{(\!K\!)}q_K}\!\prod_{m=0}^{K-1}(1-p^{(\!m\!)})}.
    \label{eq-ap1-5}
\end{equation}
By combining (\ref{eq-ap1-5}) with (\ref{eq3-2-2-1}), $\beta_{\bar{C}}$ can further be written as (\ref{eq3-2-2-2}).

\subsubsection{Derivation of $p^{(k)}$}

In the saturated condition, given that the channel is not reserved, a node's transmission is successful if and only if all the other nodes have non-capture-state HOL batches and do not transmit. We can then write the steady-state probability of successful transmission conditioned on the channel not being reserved for each HOL batch whose first packet is at the $k$-th backoff stage, $p^{(k)}$, as
\begin{equation}
    \begin{split}
        p^{(\!k\!)} \!=\! & \Pr\{\text{All the other nodes have non-capture-state} \\
        & \text{HOL batches and do not transmit}\,|\,\text{Channel is}\\
        & \text{not reserved, and the HOL batch's first packet}\\
        & \text{is in }k\text{-th backoff stage}\} \\
        = & p_1^{(k)}\cdot p_2^{(k)},
    \end{split}
    \label{eq-ap1-6}
\end{equation}
where $p_1^{(k)}=\Pr\{\text{All the other nodes do not transmit }|\text{ All}$ $\text{the other nodes have non-capture-stage HOL batches, channel}$ $\text{is not reserved, and the HOL batch's first packet is in }k\text{-th}$ \\ $\text{backoff stage}\}$ and $p_2^{(k)}=\Pr\{\text{All the other nodes have non-}$ $\text{capture-state HOL batches }|\text{ Channel is not reserved, and the}$ $\text{HOL batch's first packet is in }k\text{-th backoff stage}\}$.

For $p_1^{(k)}$, note that conditioned on no reserved transmission in the channel, whether a HOL batch's first packet is transmitted is independent of the states of other HOL batches. For a HOL batch, the probability of being at a non-capture state is $\tfrac{\tilde{\pi}_T}{M}+\sum_{l=0}^{K}\tilde{\pi}_{B_l}$ with capture-free backoff, or $\sum_{l=n_C}^{K}\tilde{\pi}_{B_l}$ with capture-based backoff. The transmission probability for a State-T or State-$B_l$ HOL batch is $q_0$ or $q_l$, respectively, $0 \leq l \leq K$. $p_1^{(k)}$ can then be written as
\begin{equation}
    p_1^{(k)} \!=\! \biggl(\!1-\tfrac{\frac{\tilde{\pi}_T}{M}q_0+\sum_{l=0}^{K}\tilde{\pi}_{B_l}q_l}{\frac{\tilde{\pi}_T}{M}+\sum_{l=0}^{K}\tilde{\pi}_{B_l}}\!\biggr)^{n-1}\!,\;p_1^{(k)} \!=\! \biggl(\!1-\tfrac{\sum_{l=n_C}^{K}\tilde{\pi}_{B_l}q_l}{\sum_{l=n_C}^{K}\tilde{\pi}_{B_l}}\!\biggr)^{n-1}\!,
    \label{eq-ap1-7}
\end{equation}
for capture-free and capture-based backoff, respectively, $0 \leq k \leq K$. By noting from (\ref{eq3-2-1-1})-(\ref{eq3-2-1-3}) that $\tfrac{\tilde{\pi}_T}{M}+\tilde{\pi}_{B_0}=\tfrac{\tilde{\pi}_T}{\beta^{(0)}q_0M}$, $\tilde{\pi}_{B_l}=\tfrac{\prod_{m=0}^{l-1}(1-p^{(m)})}{\beta^{(l)}q_l}\cdot\tfrac{\tilde{\pi}_T}{M}$ for $1\leq l \leq K-1$, and $\tilde{\pi}_{B_K}=\tfrac{\prod_{m=0}^{K-1}(1-p^{(m)})}{p^{(K)}\beta^{(K)}q_K}\cdot\tfrac{\tilde{\pi}_T}{M}$, (\ref{eq-ap1-7}) can further be written as
\begin{equation}
    p_1^{(k)} \!=\! \Biggl(\!1-\tfrac{\sum_{l=n_C}^{K-1}\tfrac{1}{\beta^{(l)}}\prod_{m=0}^{l-1}(\!1-p^{(m)}\!) + \tfrac{1}{p^{(\!K\!)}\beta^{(\!K\!)}}\prod_{m=0}^{K-1}(\!1-p^{(m)}\!)}{\sum_{l=n_C}^{K-1}\tfrac{1}{\beta^{(l)}q_l}\prod_{m=0}^{l-1}(\!1-p^{(m)}\!) + \tfrac{1}{p^{(\!K\!)}\beta^{(\!K\!)}q_K}\prod_{m=0}^{K-1}(\!1-p^{(m)}\!)}\!\Biggr)^{n-1},
    \label{eq-ap1-8}
\end{equation}
for $0 \leq k \leq K$.

For $p_2^{(k)}$, note that with capture-free backoff, all the states are non-capture states, and thus $p_2^{(k)}=1$ for $0\leq k \leq K$. With capture-based backoff, on the other hand, $p_2^{(k)}$ is dependent on the backoff stage $k$. For each HOL batch, define event $E_1^{(k)}$ as the event that its first packet is at the backoff stage $k$, event $E_3$ as the event that the channel is not reserved, and event $E_4$ as the event that there exists another capture-state HOL batch. $p_2^{(k)}$ can be written as
\begin{equation}
    p_2^{(k)} = 1- \Pr\{E_4 | E_1^{(k)}\&E_3\} = 1 - \tfrac{\Pr\{E_1^{(k)} | E_3\&E_4\}\Pr\{E_3\&E_4\}}{\Pr\{E_1^{(k)}\&E_3\}},
    \label{eq-ap1-9}
\end{equation}
for $0 \leq k \leq K$. Note that if a node's HOL batch is in a capture state, all the other nodes' HOL batches must be in non-capture states. Thus, for $0 \leq k \leq n_C-1$, we have $\Pr\{E_4 | E_1^{(k)}\&E_3\}=0$ and $p_2^{(k)}=1$. For the HOL batch whose first packet is at a non-capture state B$_k$, $n_C \leq k \leq K$, there exists a capture-state HOL batch and the channel is not reserved if and only if another node's HOL batch is at State $\text{B}_0,\cdots,\text{B}_{n_C-1}$, or the first time slot of State T. Also note that there is at most one node that has a capture-state HOL batch. $\Pr\{E_3\&E_4\}$ can then be written as
\begin{equation}
    \Pr\{E_3\&E_4\} = (n-1)\biggl(\tfrac{\tilde{\pi}_T}{M}+\sum_{l=0}^{n_C-1}\tilde{\pi}_{B_l}\biggr).
    \label{eq-ap1-10}
\end{equation}
For a node, given that another node has a capture-state HOL batch and the channel is not reserved, its HOL batch can only be at State $\text{B}_{n_C},\cdots,\text{B}_K$, and thus the probability that the first packet in its HOL batch is at the backoff stage $k$ can be written as
\begin{equation}
    \Pr\{E_1^{(k)} | E_3\&E_4\} = \tfrac{\tilde{\pi}_{B_k}\beta^{(k)}}{\sum_{l=n_C}^{K}\tilde{\pi}_{B_l}\beta^{(l)}},
    \label{eq-ap1-11}
\end{equation}
for $n_C \leq k \leq K$. The probability that the HOL batch's first packet is at the backoff stage $k$ and the channel is not reserved $\Pr\{E_1^{(k)}\&E_3\}$ is given by
\begin{equation}
    \Pr\{E_1^{(k)}\&E_3\} = \tilde{\pi}_{B_k}\beta^{(k)},
    \label{eq-ap1-12}
\end{equation}
for $n_C \leq k \leq K$. By substituting (\ref{eq-ap1-10})-(\ref{eq-ap1-12}) into (\ref{eq-ap1-9}), $p_2^{(k)}$ for capture-based backoff can then be derived as
\begin{equation}
    p_2^{(k)} =
    \begin{cases}
        1 & 0\!\leq\! k \!\leq\! n_C\!-\!1 \\
        1-\tfrac{(n-1)\bigl(\frac{\tilde{\pi}_T}{M}+\sum_{l=0}^{n_C-1}\tilde{\pi}_{B_l}\bigr)}{\sum_{l=n_C}^{K}\tilde{\pi}_{B_l}\beta^{(l)}} & n_C \!\leq\! k \!\leq\! K.
    \end{cases}
    \label{eq-ap1-13}
\end{equation}
By combining $p_2^{(k)}=1$ for capture-free backoff and (\ref{eq-ap1-13}) for capture-based backoff with (\ref{eq3-2-1-1})-(\ref{eq3-2-1-3}), we have
\begin{equation}
    p_2^{(k)} =
    \begin{cases}
        1 \hspace{4.8cm} 0\leq k \leq n_C-1 \\
        1-\tfrac{(n-1)\sum_{l=0}^{n_C-1}\tfrac{1}{\beta^{(l)}}\prod_{m=0}^{l-1}(1-p^{(m)})}{\sum_{l=n_C}^{K-1}\tfrac{1}{q_l}\prod_{m=0}^{l-1}(1-p^{(m)}) + \tfrac{1}{p^{(K)}q_K}\prod_{m=0}^{K-1}(1-p^{(m)})} \\
        \hspace{5.3cm} n_C \leq k \leq K.
    \end{cases}
    \label{eq-ap1-14}
\end{equation}

By combining (\ref{eq-ap1-6}), (\ref{eq-ap1-8}), (\ref{eq-ap1-14}) with (\ref{eq3-2-2-1})-(\ref{eq3-2-2-2}), the steady-state probability of successful transmission conditioned on the channel not being reserved $p_C$ for each capture-state HOL batch and $p_{\bar{C}}$ for each non-capture-state HOL batch can be obtained as (\ref{eq3-2-2-3})-(\ref{eq3-2-2-4}).

\vspace{-0.5cm}
\section{Derivation of (\ref{eq4-4-1})}\label{sec_ap3}

Let $\bar{\lambda}_{out,i}^T$ and $\sigma_{\lambda_{out,i}^T}^2$ denote the mean and the variance of\vspace{0.1cm} $\lambda_{out,i}^T$, i.e., the output rate of Node $i$'s data queue by time $T$, respectively, $i\in\mathcal{N}$. Given that all the nodes have the same parameter setting, they have identically distributed $\lambda_{out,i}^T$ with $\bar{\lambda}_{out,i}^T=\bar{\lambda}_{out}^T$ and $\sigma_{\lambda_{out,i}^T}^2=\sigma_{\lambda_{out}^T}^2$ for all $i\in\mathcal{N}$. The short-term fairness index $J_T$ can further be written from (\ref{eq1-4}) as
\begin{equation}
    J_T = \tfrac{(\frac{1}{n}\sum_{i=1}^{n}\lambda_{out,i}^T)^2}{\frac{1}{n}\cdot\sum_{i=1}^{n}(\lambda_{out,i}^T)^2} \overset {n\gg 1, T\gg 1}{\approx} \tfrac{(\bar{\lambda}_{out}^T)^2}{(\bar{\lambda}_{out}^T)^2+\sigma_{\lambda_{out}^T}^2}.
    \label{eq-ap3-1}
\end{equation}

To derive $\bar{\lambda}_{out}^T$ and $\sigma_{\lambda_{out}^T}^2$, note that the output rate of Node $i$'s data queue by time $T$ can be written from (\ref{eq1-1}) as
\begin{equation}
    \lambda_{out, i}^T \overset {T\gg M}{\approx} \tfrac{M\cdot R_{i,T}}{T},
    \label{eq-ap3-2}
\end{equation}
where $R_{i,T}$ denotes the number of successfully transmitted data batches by time $T$ of Node $i$, $i\in\mathcal{N}$. When Node $i$'s queue is saturated, $R_{i,T}$ can be modeled as a renewal process:
\begin{equation}
    R_{i,T} = \max \Bigl\{m: \sum_{j=1}^m D_{i,j}\leq T\Bigr\},
    \label{eq-ap3-3}
\end{equation}
where $D_{i, j}$ denotes the service time of batch $j$ of Node $i$'s queue. As $T\to \infty$, $R_{i,T}\to\mathcal{N}\Bigl(\tfrac{T}{\bar{D}_i}, \tfrac{\sigma_{D_i}^2}{\bar{D}_i^3}T\Bigr)$ \cite{gallager2012discrete}, where $\bar{D}_i$ and $\sigma_{D_i}^2$ denote the mean and variance of $D_{i,j}$, respectively. Given that all the nodes have the same parameter setting, they have identically distributed service times of data batches with $\bar{D}_i=\bar{D}$ and $\sigma_{D_i}^2=\sigma_{D}^2$ for all $i\in\mathcal{N}$. By combining $R_{i,T}\to\mathcal{N}\Bigl(\tfrac{T}{\bar{D}}, \tfrac{\sigma_{D}^2}{\bar{D}^3}T\Bigr)$ with (\ref{eq-ap3-2}), $\{\lambda_{out, i}^T\}_{i\in\mathcal{N}}$ can then be approximated as i.i.d. Gaussian random variables with the mean and variance:
\begin{equation}
    \bar{\lambda}_{out}^T = \tfrac{M}{\bar{D}},\;\;\;\sigma_{\lambda_{out}^T}^2=\tfrac{M^2 \sigma_{D}^2}{\bar{D}^3 T}.
    \label{eq-ap3-4}
\end{equation}
By substituting (\ref{eq-ap3-4}) into (\ref{eq-ap3-1}), the short-term fairness index can be derived as (\ref{eq4-4-1}).

\section{Derivation of (\ref{eq4-4-3})-(\ref{eq4-4-4})}\label{sec_ap4}

$G_{D}^{\prime}(1)$ and $G_{D}^{\prime\prime}(1)$ can be obtained based on the Markov renewal process of HOL batches established in Section \ref{sec4-1}. Specifically, according to Fig. \ref{fig3-2-2}, we have
\begin{equation}
    D =
    \begin{cases}
        Y_T & \text{with probability } p^{(0)}\beta^{(0)}q_0 \\
        D_{B_0} & \text{with probability } 1-\beta^{(0)}q_0 \\
        D_{B_1} & \text{with probability } \beta^{(0)}q_0(1-p^{(0)}),
    \end{cases}
    \label{eq-ap4-1}
\end{equation}
\begin{equation}
    D_{B_k} =
    \begin{cases}
        Y_{B_k}+Y_T & \text{with probability } p^{(k)} \\
        Y_{B_k}+D_{B_{k+1}} & \text{with probability } 1-p^{(k)},
    \end{cases}
    \label{eq-ap4-2}
\end{equation}
for $0 \leq k \leq K-1$, and $D_{B_K} = Y_{B_K}+Y_T$, where $D_u$ denotes the time of a HOL batch spent from the beginning of State $u$ till the service completion, and $Y_u$ denotes the sojourn time of a HOL batch at State $u$, $u\in\mathcal{S}$. Note that $Y_T=M$, and $Y_{B_k}$ can be regarded as a geometrically distributed random variable with parameter $\beta^{(k)}q_k$ for $0 \leq k \leq K-1$, and $p^{(K)}\beta^{(K)}q_K$ for $k=K$. Based on (\ref{eq-ap4-1})-(\ref{eq-ap4-2}), $G_{D}^{\prime}(1)$ and $G_{D}^{\prime\prime}(1)$ can be obtained as (\ref{eq4-4-3}) and (\ref{eq4-4-4}), respectively.

\balance

\normalem
\bibliographystyle{IEEEtran}
\bibliography{IEEEabrv, ref}

\end{document}